# Two-Sided EWMA Charts for Monitoring Double Bounded Processes

Argyro Lafatzi[1] and Athanasios C. Rakitzis[2,*]

[1]Department of Statistics and Actuarial-Financial Mathematics, University of the Aegean, Karlovasi, Samos, Greece.

[2]Department of Statistics and Insurance Science, University of Piraeus, Piraeus, Greece.

**Abstract**

In this work, we study the performance of two-sided EWMA charts for monitoring double bounded processes using individual observations. Specifically, the term double bounded refers to observations in the interval (0, 1) and thus, these charts are suitable for monitoring rates, proportions and percentages. There are several models that can be used to describe this kind of data (and the respective processes, as well) such as the Beta distribution, the Simplex distribution and the Unit Gamma distribution. For each of these three models, we provide the statistical design and the performance of the proposed EWMA charts. Also, apart from providing the appropriate values for the design parameters of each chart, we investigate how much the performance of the EWMA schemes is affected by using the values of control limits which have not been calculated under the true model. Finally, an illustrative example is also presented.

## 1. Introduction

Statistical process monitoring (SPM) is a collection of tools that allows the monitoring of a process. Among them, the control chart is the most widely used SPM tool. Control charts have been proposed by Walter A. Shewhart (1926) in the 1920's. Usually, they are used in industry and in the monitoring of a manufacturing process, in order to detect any abnormal (usually unwanted) situation which affects production process. In case of the presence of these unwanted situations, the quality of the produced items deteriorates, and the percentage of defective items increases. Nowadays, applications of control charts can be found in several areas of applied science, such as healthcare, environmental monitoring, social networks and big data analysis (see, for example, Woodall (2006), Woodall et al. (2017), Bersimis et al. (2018), Aykroyd et al. (2019)).

[*] Corresponding author: arakitz@unipi.gr

In several cases, an item (or a product) is classified as defective (or nonconforming) if at least one quality characteristic does not satisfy the specifications that have been set during its design phase. With the term fraction nonconforming items we refer to the number of nonconforming items over the total number of items (or products), in say, a sample or a lot. Let us denote as $p$ the proportion of nonconforming items that are produced by a process and let also $n$ be the size of a sample that it is collected from the process. If we denote as $X_{ij}$ a random variable (r.v.) that takes the value 1 if a product is classified as non-conforming or the value 0, otherwise, then $X_{ij} \sim B(1,p)$ and $X_i = \sum_{j=1}^{n} X_{ij}$ is the r.v. that denotes the total number of non-conforming items, in the sample. Consequently, $X_i \sim B(n,p)$, $i = 1,2,....$, where $B(n,p)$ denotes the binomial distribution with parameters $n \in \{1,2,...\}$ and $p \in (0,1)$. Therefore, the data that are available under this setup are referred as "results from Bernoulli experiments" or, in general, as attributes data, since it is not possible to obtain a numerical value from the characteristic that describes the quality of the produced items, but we rather record whether an attribute is present or not in the item. For example, in health-related processes we record whether a patient is a smoker or not, if it has been infected or not after a surgical operation etc.

Control charts for attributes data are widely used and two of the most common attributes control charts are the $p$ and $np$ charts (see, for example Montgomery (2013)), which are used for monitoring a process by monitoring the proportion or the number of non-conforming items, respectively. Especially, in the case of the $p$ chart, the values of the points plotted on the chart are values, in general, in the interval [0, 1].

However, there are cases where the values of a quality characteristic $X$ are (in general) in the interval [0, 1] but they are not results from Bernoulli experiments. For example, the monthly percentage of unemployment or the inflation in a country, the daily relative humidity in a city, the percentage of fat in the body of a patient or the percentage of a specific ingredient in a food or a pharmaceutical product. The common term for such values is continuous proportions. When we are interested in monitoring continuous proportions, the usual $p$ and $np$ charts cannot be applied, and alternative monitoring schemes need to be established and used. In the recent years, there is an increased interest in providing models and process monitoring techniques for data that are doubly bounded, e.g. in [0, 1] or (0, 1). This work focuses mainly on the monitoring of continuous proportions.

A well-known model for modelling a double bounded process is that of Beta distribution. It is a flexible continuous distribution that can model a large variety of data in (0, 1), having different shapes (Kieschnick and McCullough (2003)). Gupta and Nadarajah (2004) presented several applications of the Beta distribution and it seems that these authors are the first who considered an application using control charts. Sant'Anna and ten Caten (2012) developed and applied Shewhart control charts based on the Beta distribution to monitor fraction data, as an alternative method for monitoring proportions (instead of the $p$ chart). Bayer et al. (2018) studied the beta regression chart, when the proportion of non-conforming depends on several characteristics (which can be considered as independent variables) and a linear regression model can be established to describe this relation. Ho et al. (2019) studied Shewhart control charts under three different probability models (Beta, Simplex and Unit Gamma) for double bounded processes with values in (0, 1). Lima-Filho and Bayer (2021) proposed and studied a Shewhart chart based on the Kumaraswamy distribution (see Kumaraswamy (1980)) while a median control chart based on the Kumaraswamy and the unit-Weibull distributions was studied by Lima-Filho et al. (2020).

In all the previously mentioned works, the proposed charts are Shewhart-type charts. It is well-known that for small and/or moderate shifts in process parameter(s), the Shewhart charts are not sensitive enough and they cannot detect them quickly. This is attributed to the fact that in a Shewhart chart the decision whether a process is in-control or out-of-control is based only on the most recent observation (e.g. on the most recent available proportion/percentage). Control charts with memory, such as the Cumulative Sum (CUSUM, Page (1954)) and the Exponentially Weighted Moving Average (EWMA, Roberts (1959)) charts can be used, offering improved sensitivity in the detection of small and moderate shifts. Both CUSUM and EWMA charts are control charts with memory and the values that are plotted on each chart, incorporate information from both recent and past values. Therefore, they have an increased sensitivity in the detection of shifts of small and moderate magnitude in process parameters.

In this work, in order to provide an improved method in monitoring a double bounded process, we propose and study two-sided EWMA charts for individual observations in (0,1). Also, motivated by the work of Ho et al. (2019) we investigate how much the performance of an EWMA chart is affected when it is designed under a different model rather than the correct one. According to Montgomery (2013), page 439, an EWMA chart for individual observations that is well designed can be considered as a viable nonparametric (distribution-free) procedure, particularly for Phase II monitoring, in a wide range of applications. Since there are several different models for double bounded processes in (0,1) is it also interesting to investigate under which circumstances this distribution-free property maintains.

The outline of the paper is the following. In Section 2, we provide in brief the properties of the three continuous distributions, Beta, Simplex and Unit Gamma, which are used as possible models for the process. In Section 3, we present the EWMA charts based on the three above mentioned distributions, along with the algorithmic procedure for the statistical design of the charts as well as for calculating the necessary performance measures for its out-of-control performance. In Section 4, we provide the results of an extensive simulation study, regarding the performance of the two-sided EWMA charts. Also, we present comparisons between the proposed EWMA charts and the Shewhart charts studied by Ho et al. (2019). A numerical example is given in Section 5. Finally, in Section 6, we provide conclusions, practical guidelines, and topics for future research.

## 2. Distributions for Modelling Continuous Proportions

Following the work of Ho et al. (2019), we consider three continuous probability models with bounded support and specifically, the (0, 1) interval. Next, we present in brief their properties.

### 2.1 Beta Distribution

Let $X$ be a r.v. which follows a Beta distribution with parameters $\alpha > 0$, $\beta > 0$ (i.e. $X \sim Beta(\alpha, \beta)$). Then, its probability density function (p.d.f.) is given by

$$f_{Beta}(x|\alpha, \beta) = \frac{x^{\alpha-1}(1-x)^{\beta-1}}{B(\alpha, \beta)}, \ x \in (0,1),$$

where $B(\alpha, \beta) = \Gamma(\alpha)\Gamma(\beta)/\Gamma(\alpha+\beta)$ is the Beta function and $\Gamma(y) = \int_0^\infty t^{y-1}e^{-t}dt$ is the Gamma function. Also, its cumulative distribution function (c.d.f.) is given by

$$F_{Beta}(x|\alpha, \beta) = \frac{B_x(\alpha, \beta)}{B(\alpha, \beta)},$$

where $B_x(\alpha, \beta) = \int_0^x t^{\alpha-1}(1-t)^{\beta-1}dt$ is the incomplete Beta function. The expected value and the variance of the $Beta(\alpha, \beta)$ distribution equal

$$E(X) = \frac{\alpha}{\alpha+\beta} \text{ and } V(X) = \frac{\alpha\beta}{(\alpha+\beta)^2(\alpha+\beta+1)}.$$

In this work we use the re-parametrized Beta distribution (see Ferrari and Cribari-Neto (2004)) which arises if we set $\alpha = \mu\phi$ and $\beta = (1-\mu)\phi$, for $\mu \in (0,1)$ and $\phi > 0$. Therefore, the p.d.f. is given by

$$f_{Beta}(x|\mu,\phi) = \frac{x^{\mu\phi-1}(1-x)^{(1-\mu)\phi-1}}{B(\mu\phi,(1-\mu)\phi)}, \quad x \in (0,1),$$

while the expected value and the variance are now equal to

$$E(X) = \mu, \quad V(X) = \frac{\mu(1-\mu)}{\phi+1}. \tag{1}$$

Parameter $\phi$ can be viewed as a precision (or dispersion) parameter; when $\mu$ remains unchanged, if $\phi$ increases then the variance of the $Beta(\mu,\phi)$ distribution decreases.

## 2.2 Simplex Distribution

Let $X$ be a r.v. that follows a Simplex distribution with parameters $\mu \in (0,1)$, $\sigma^2 > 0$ (i.e., $X \sim S(\mu,\sigma^2)$). Then, its p.d.f. is given by (see, for example Jørgensen (1997))

$$f_{Simplex}(x|\mu,\sigma^2) = \frac{1}{\sqrt{2\pi\sigma^2 x^3(1-x)^3}} e^{\left(-\frac{1}{2\sigma^2}d(x;\mu)\right)}, \quad x \in (0,1),$$

where the term $d(x;\mu)$ is known as deviance function and it is given by

$$d(x;\mu) = \frac{(x-\mu)^2}{x(1-x)\mu^2(1-\mu)^2}.$$

The expected value and the variance of the $S(\mu,\sigma^2)$ distribution equal

$$E(X) = \mu, \quad V(X) = \frac{1}{\sqrt{2\sigma^2}} e^{\left(\frac{1}{2\sigma^2\mu^2(1-\mu)^2}\right)} \Gamma\left(\frac{1}{2}, \frac{1}{2\sigma^2\mu^2(1-\mu)^2}\right), \tag{2}$$

where $\Gamma(r,s) = \int_s^\infty u^{r-1} e^{-u} du$ is the incomplete Gamma function.

## 2.3 Unit Gamma Distribution

Let $X$ be a r.v. that follows a Unit Gamma distribution with parameters $\theta > 0$, $\tau > 0$ (i.e., $X \sim uGA(\theta,\tau)$). Then, its p.d.f. is given by (see also Grassia (1977))

$$f_{uGA}(x|\theta,\tau) = \frac{\theta^\tau}{\Gamma(\tau)} x^{\theta-1} \left(\log\left(\frac{1}{x}\right)\right)^{\tau-1}, \quad x \in (0,1),$$

while its expected value and variance are equal to

$$E(X) = \left(\frac{\theta}{\theta+1}\right)^{\tau}, \quad V(X) = \left(\frac{\theta}{\theta+2}\right)^{\tau} - \left(\frac{\theta}{\theta+1}\right)^{2\tau}.$$

In this work, we consider the re-parametrized Unit Gamma distribution (see Mousa et al. (2016)) which arises by setting $\theta = \frac{\mu^{1/\tau}}{(1-\mu^{1/\tau})}$. Then, the p.d.f. takes the following form

$$f_{uGA}(x|\mu,\tau) = \frac{\left(\frac{\mu^{1/\tau}}{(1-\mu^{1/\tau})}\right)^{\tau}}{\Gamma(\tau)} x^{\frac{\mu^{1/\tau}}{(1-\mu^{1/\tau})}-1} \left(\log\left(\frac{1}{x}\right)\right)^{\tau-1}, \quad x \in (0,1),$$

where $\mu \in (0,1)$, $\tau > 0$. Therefore, under the re-parametrized Unit Gamma model, the expected value and the variance equal

$$E(X) = \mu, \quad V(X) = \mu\left(\frac{1}{(2-\mu^{1/\tau})^{\tau}} - \mu\right). \tag{3}$$

## 3. EWMA Control Charts for Double Bounded Processes

In this section, we present the EWMA control charts for monitoring a double bounded process with individual observations. This means that at each sampling stage a single value is available. First, when the process is in-control (IC) we denote its process mean level as $\mu_{0,X} = \mu_0$ which is the IC proportion. Also, the precision (or dispersion) parameters for each model are denoted as $\phi_0$ (for the Beta model), $\sigma_0$ (for the Simplex model) and $\tau_0$ (for the Unit Gamma model). Therefore, the IC process variability is denoted as $\sigma_{0,X}$, which is evaluated from the respective formulas for the $V(X)$ (see equations (1)-(3)), by substituting the IC values of each model. Also, we assume that the IC values of the process parameters are known, and the proposed charts are suitable for monitoring the process in real-time, or for a Phase II analysis, as it is sometimes called the real-time monitoring of a process.

When the process is out-of-control (OOC), we assume that the presence of assignable causes affects only the process mean level, which changes from $\mu_0$ to $\mu_1 \neq \mu_0$, with $\mu_1 \in (0,1)$. Specifically, when $\mu_1 > \mu_0$ an increasing shift has occurred while for $\mu_1 < \mu_0$, the process mean level has been decreased. Note also that in this work, we assume that the value of the dispersion parameter remains unaffected by the presence of assignable causes in the process. The aim is to detect changes in the IC process mean level which equals $\mu_0$ and is not directly affected by changes in the parameters $\phi$, $\sigma$ and $\tau$.

Let us consider an EWMA control chart with two control limits, an upper control limit ($UCL$) and a lower control limit ($LCL$). On an EWMA control chart, the values of the following statistic

$$Z_t = \lambda X_t + (1-\lambda)Z_{t-1},$$

are plotted, where $X_t$ is the single value (case of individual observations) that it is collected (or recorded) at the $t$-th sampling stage ($t = 1, 2, ...$). Also, the initial value $Z_0 = \mu_0$, i.e., it equals the IC expected value of $X$. The smoothing parameter $\lambda \in (0,1]$ controls the weight that it is given to the recent observations from the process. Therefore, small values give less weight on the recent observations and more weight on the past. The case is reversed for larger $\lambda$ values. According to Montgomery (2013), common values for $\lambda$ are in the interval [0.05, 0.25]. For $\lambda = 1$, the two-sided EWMA chart coincides with a two-sided Shewhart chart, since $Z_t = X_t$.

In this work, we assume that the control limits $LCL$, $UCL$ and the center line $CL$ of the two-sided EWMA chart are given by (see also Montgomery (2013), Borror et al. (1999), Human et al. (2011))

$$LCL = \mu_{Z_t} - L\sigma_{Z_t} = \mu_{0,X} - L\sigma_{0,X}\sqrt{\frac{\lambda}{2-\lambda}}, \quad UCL = \mu_{Z_t} + L\sigma_{Z_t} = \mu_{0,X} + L\sigma_{0,X}\sqrt{\frac{\lambda}{2-\lambda}},$$

$$CL = \mu_{Z_t} = \mu_{0,X},$$

where $\mu_{Z_t} = E(Z_t)$ and $\sigma_{Z_t}^2 = V(Z_t)$.

The above limits are also known as steady-state control limits for the two-sided EWMA chart. The chart gives an out-of-control signal at the $t$-th sampling stage if $Z_t \notin [LCL, UCL]$. The number of points plotted on the chart until it gives for the first time an OOC signal is known as run length ($RL$) and it is a r.v. The distribution of the $RL$ (also known as run-length distribution) is mainly used for evaluating the performance of the chart. The most common performance measure is its expected value $E(RL)$ or the average run length ($ARL$) while the standard deviation of the run-length ($SDRL = \sqrt{V(RL)}$) and the median of the run length distribution (median run length or $MRL$), are also used.

In the case of a two-sided Shewhart control chart, the $RL$ is a geometric r.v. with parameter $p_{out}$ which is the probability for a point to be outside the interval $[LCL_{SH}, UCL_{SH}]$, where $LCL_{SH}, UCL_{SH}$ are the control limits of the two-sided Shewhart chart. Therefore, the $ARL$, $SDRL$ and $MRL$ are given by

$$ARL = \frac{1}{p_{out}}, \quad SDRL = \frac{\sqrt{1-p_{out}}}{p_{out}}, \quad MRL = \left\lceil \frac{\log(0.5)}{\log(1-p_{out})} \right\rceil.$$

In order to apply a two-sided EWMA chart in the monitoring of a double bounded process, except for the IC values of the process parameters, the values for $\lambda$ and $L$ need to be selected. Thus, the $RL$ distribution of the two-sided EWMA needs to be determined. It is also known that in the case of a two-sided EWMA chart, the $RL$ distribution is not a geometric one. In this work, we apply the method of Monte Carlo simulation to determine $L$ for a given $\lambda$ value, so as the performance of the chart is the desired one. Specifically, the steps of the algorithmic procedure for the determination of $(\lambda, L)$ are given in Table 1 (see also Alevizakos and Koukouvinos (2020)).

**Table 1.** Steps for Determining $L$ via Monte Carlo Simulation.

| |
|---|
| 1. Choose the IC value for the process parameters. These are $(\mu_0, \phi_0)$ for the Beta distribution, $(\mu_0, \sigma_0)$ for the Simplex distribution and $(\mu_0, \tau_0)$ for the Unit Gamma distribution. |
| 2. Choose the desired IC $ARL$ value, say $ARL_0$ and prespecify the value for $\lambda$. |

3. Use a starting value $L = 0.001$ and calculate the steady-state control limits of the two-sided EWMA chart.

4. Simulate 10000 IC processes with the specific IC process parameters from Step 1 and for each sequence, record the number of values until the first false alarm is triggered.

5. Estimate the IC $ARL$ as the sample mean of the 10000 $RL$ values obtained in Step 4. If $ARL \notin (ARL0 - \xi, ARL0 + \xi)$, where $\xi$ is a tolerance number, increase $L$ by 0.001 and go back to Step 4. Otherwise, go to Step 6.

6. Use the value for $L$ that has been obtained in the previous step, set up the control limits for the two-sided EWMA chart and declare the process as OOC at sample $t$, if $Z_t \notin [LCL, UCL]$.

Once the value of $L$ has been determined, then we can proceed with the estimation of $ARL$ (and $SDRL$, $MRL$) when the process is OOC, for a given shift in $\mu_0$. The steps of the algorithmic procedure are given in Table 2.

**Table 2.** Steps for Estimating OOC $ARL$ via Monte Carlo Simulation.

1. Choose the IC value for the process parameters. These are $(\mu_0, \phi_0)$ for the Beta distribution, $(\mu_0, \sigma_0)$ for the Simplex distribution and $(\mu_0, \tau_0)$ for the Unit Gamma distribution.

2. Setup a two-sided EWMA control chart using the values $(\lambda, L)$ obtained during the design phase of the chart.

3. Choose the shift in $\mu_0$, or equivalently, the OOC value $\mu_1$.

4. Simulate 10000 OOC processes with parameters $(\mu_1, \phi_0)$ (for the Beta distribution) or $(\mu_1, \sigma_0)$ (for the Simplex distribution) or $(\mu_1, \tau_0)$ (for the Unit Gamma distribution).

> 5. For each of the 10000 sequences, record the number of values until the first (true) alarm is triggered.
> 6. Estimate the OOC $ARL$ as the sample mean of the 10000 $RL$ values from Step 5 as well as the OOC $SDRL$ and the OOC $MRL$.

## 4. Numerical Results

In this section, we present the results of an extensive numerical study regarding the performance and the design of the proposed EWMA charts. The aim is dual: First we aim to compare the EWMA charts with the Shewhart charts for rates and proportions that have been studied by Ho et al. (2019). And then, to assess how much the EWMA charts based on the Beta, Simplex and Unit Gamma distribution are affected when a different model (different than the true one) is used for the calculation of the control limits of the chart. It should be mentioned that our intention is not to provide a guidance on the probability model that it is less affected when 'wrong' control limits are used but to investigate how much the EWMA chart (for various $\lambda$ values) is affected by the use of 'wrong' limits.

First, as already said, we assume that the presence of assignable causes affects only the mean of the process; that is, when the process is OOC, the expected value $\mu$ shifts from $\mu_0$ to a value $\mu_1 \neq \mu_0$. When two-sided charts are used, we are interested in detecting shifts either increasing or decreasing. In this work, we assume that $\mu_0 = 0.2$ and the OOC mean is $\mu_1 = \mu_0 \pm \delta$, with $\delta \in \{0.02, 0.04, 0.06, 0.08\}$. Following Ho et al. (2019), we consider four different levels of dispersion (different values for the dispersion parameters in each of the three models) which remain unchanged in the presence of assignable causes. These values are summarized in Table 3, which consists of the coefficient of variation (CV), the skewness coefficient and the kurtosis coefficient (columns 'Skew' and 'Kurt'). The three models for each of the Cases 1-4 have similar properties. This can be also deduced by their p.d.f. which are given in Figure 1.

**Table 3.** Properties of the different models with $\mu_0 = 0.2$, Cases 1-4.

| | Beta | | | | |
|---|---|---|---|---|---|
| Case | $\phi$ | $\sigma_{0,X}$ | CV | Skew | Kurt |
| 1 | 290 | 0.02344842 | 0.1172421 | 0.1575075 | 2.943123 |
| 2 | 148 | 0.03276928 | 0.1638464 | 0.2442577 | 3.048881 |
| 3 | 80 | 0.04444444 | 0.2222222 | 0.3292683 | 3.088377 |
| 4 | 31 | 0.07071068 | 0.3535534 | 0.5142594 | 3.208556 |
| | Simplex | | | | |
| Case | $\sigma$ | $\sigma_{0,X}$ | CV | Skew | Kurt |
| 1 | 0.37 | 0.02355733 | 0.1177866 | 0.2558005 | 2.962745 |
| 2 | 0.50 | 0.03170082 | 0.1585041 | 0.3485029 | 3.127102 |
| 3 | 0.71 | 0.04460488 | 0.2230244 | 0.4789127 | 3.229076 |
| 4 | 1.20 | 0.07309293 | 0.3654647 | 0.7332117 | 3.486262 |
| | Unit Gamma | | | | |
| Case | $\tau$ | $\sigma_{0,X}$ | CV | Skew | Kurt |
| 1 | 155 | 0.02582828 | 0.1291414 | 0.2260785 | 3.016697 |
| 2 | 96 | 0.03279827 | 0.1639913 | 0.2852494 | 3.088033 |
| 3 | 51 | 0.04493217 | 0.2246609 | 0.38687 | 3.159441 |
| 4 | 20 | 0.07138937 | 0.3569468 | 0.5963744 | 3.360247 |

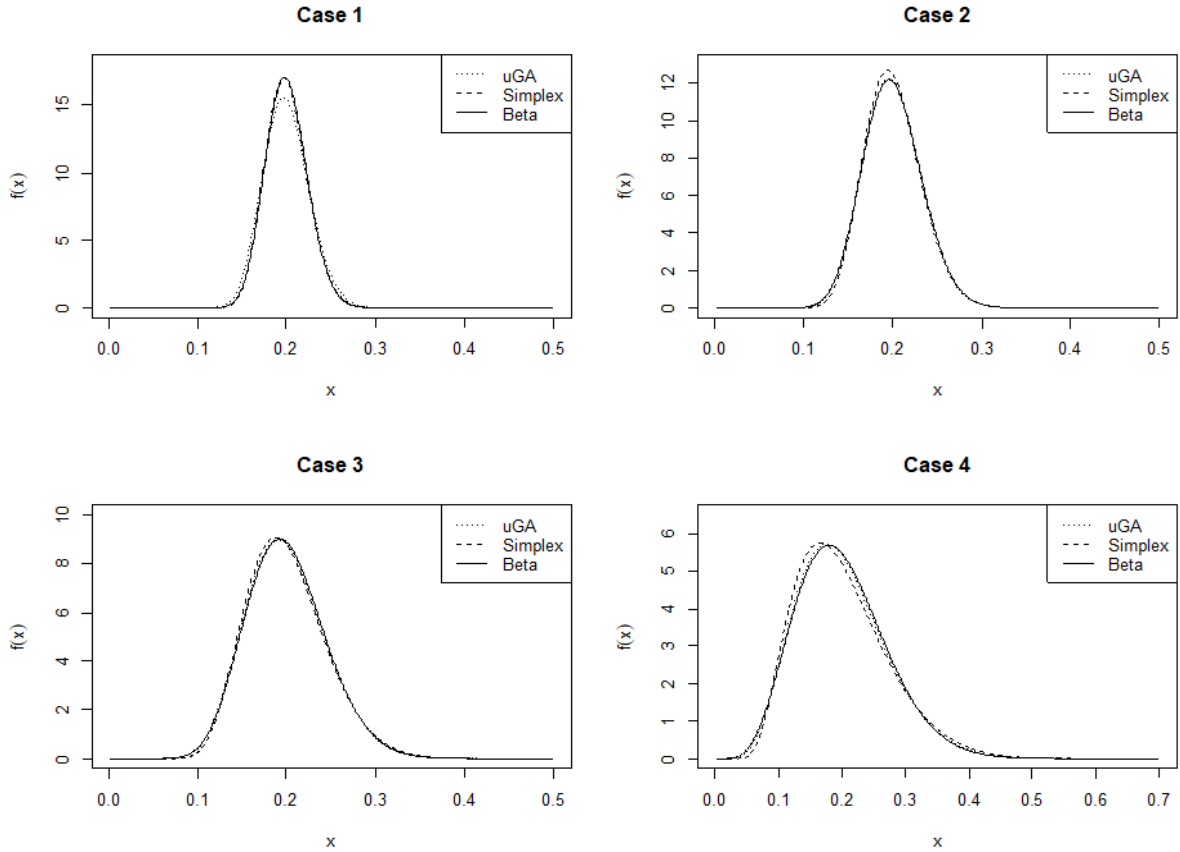

**Figure 1:** Probability density function for the Beta, Simplex and Unit Gamma distributions, Cases 1-4.

Next, we present the $ARL$ performance of the two-sided Shewhart and EWMA charts, for each of the three distributions that are used to model the double bounded process (Beta, Simplex and Unit Gamma). The false alarm rate for each chart equals $a = 0.0027$ and thus the $ARL_0$ value is approximately equal to 370.4. For $\lambda$ we used three common choices, 0.05, 0.10 and 0.20 and via Monte Carlo simulation we determined the $L$ value in order to have the desired IC performance. We applied the steps of the algorithmic procedure in Table 1 and we determined the pair of values $(\lambda, L)$. The $L$ values are given in Table A1 in the Appendix.

For the Shewhart chart we used equal tail probability limits, based on the IC distribution of individual observations $X_t$. For example, in case of Beta distribution, the limits are given by

$$LCL_{SH} = F_{Beta}^{-1}(0.00135|\mu_0, \phi_0), \quad UCL_{SH} = F_{Beta}^{-1}(0.99865|\mu_0, \phi_0),$$

where $F_{Beta}^{-1}(..|\mu_0,\phi_0)$ is the inverse c.d.f. of the $Beta(\mu_0,\phi_0)$. Similarly, for the Simplex and the Unit Gamma Shewhart charts, we only need to replace the above inverse c.d.f. with $F_{Simplex}^{-1}(...|\mu_0,\sigma_0)$ and $F_{uGA}^{-1}(...|\mu_0,\tau_0)$.

Then, for all the OOC $\mu_1$ values, we evaluated the OOC $ARL$ performance of the charts. For the two-sided EWMA charts we applied the steps of the algorithmic procedure in Table 2. The results are given in Tables 4-6. In the boldfaced entries it is the minimum $ARL$ value among the different charts.

Table 4: *ARL* performance for the two-sided Shewhart and EWMA charts, Beta model.

| $\phi_0$ | $\mu_1$ | SH | $\lambda = 0.05$ | $\lambda = 0.10$ | $\lambda = 0.20$ | $\phi_0$ | $\mu_1$ | SH | $\lambda = 0.05$ | $\lambda = 0.10$ | $\lambda = 0.20$ |
|---|---|---|---|---|---|---|---|---|---|---|---|
| 290 | 0.12 | **1.26** | 3.98 | 3.40 | 3.06 | 80 | 0.12 | 5.11 | 6.49 | 5.62 | **5.05** |
| | 0.14 | **2.34** | 4.87 | 4.23 | 3.67 | | 0.14 | 12.99 | 8.54 | 7.55 | **7.04** |
| | 0.16 | 8.04 | 6.81 | 5.96 | **5.32** | | 0.16 | 40.85 | 13.13 | **12.15** | 13.19 |
| | 0.18 | 54.60 | 13.89 | **13.20** | 14.10 | | 0.18 | 150.86 | **31.81** | 35.86 | 55.18 |
| | 0.20 | 370.37 | 370.14 | 370.30 | 370.88 | | 0.20 | 370.37 | 370.48 | 370.49 | 370.79 |
| | 0.22 | 69.71 | 14.04 | **13.29** | 13.89 | | 0.22 | 195.11 | **31.70** | 33.62 | 40.80 |
| | 0.24 | 12.26 | 6.87 | 6.06 | **5.42** | | 0.24 | 67.47 | 13.28 | **12.15** | 12.46 |
| | 0.26 | **3.71** | 4.90 | 4.25 | 3.73 | | 0.26 | 26.33 | 8.65 | 7.62 | **7.20** |
| | 0.28 | **1.78** | 3.97 | 3.46 | 3.06 | | 0.28 | 11.93 | 6.58 | 5.77 | **5.16** |
| | UCL | 0.2755 | 0.2093 | 0.2145 | 0.2224 | | UCL | 0.3506 | 0.2177 | 0.2275 | 0.2425 |
| | LCL | 0.1355 | 0.1907 | 0.1855 | 0.1776 | | LCL | 0.0884 | 0.1823 | 0.1725 | 0.1575 |
| $\phi_0$ | $\mu_1$ | SH | $\lambda = 0.05$ | $\lambda = 0.10$ | $\lambda = 0.20$ | $\phi_0$ | $\mu_1$ | SH | $\lambda = 0.05$ | $\lambda = 0.10$ | $\lambda = 0.20$ |
| 148 | 0.12 | **2.36** | 5.04 | 4.35 | 3.81 | 31 | 0.12 | 15.68 | 10.15 | **9.11** | 9.21 |
| | 0.14 | 5.72 | 6.42 | 5.52 | **4.94** | | 0.14 | 36.30 | 13.93 | **13.28** | 15.30 |
| | 0.16 | 20.11 | 9.46 | 8.40 | **8.11** | | 0.16 | 90.21 | **23.24** | 24.96 | 37.61 |
| | 0.18 | 100.51 | **21.09** | 21.40 | 27.07 | | 0.18 | 220.61 | **64.30** | 87.98 | 188.96 |
| | 0.20 | 370.37 | 370.14 | 370.44 | 370.32 | | 0.20 | 370.37 | 370.50 | 370.66 | 370.16 |
| | 0.22 | 129.21 | **21.05** | 21.19 | 24.19 | | 0.22 | 289.82 | **59.48** | 66.36 | 78.97 |
| | 0.24 | 32.24 | 9.55 | 8.47 | **8.12** | | 0.24 | 155.75 | 23.13 | **23.07** | 26.48 |
| | 0.26 | 10.61 | 6.47 | 5.64 | **5.09** | | 0.26 | 81.39 | 14.17 | **13.30** | 13.57 |
| | 0.28 | 4.53 | 5.09 | 4.41 | **3.88** | | 0.28 | 44.63 | 10.37 | 9.30 | **9.02** |
| | UCL | 0.3081 | 0.2130 | 0.2202 | 0.2313 | | UCL | 0.4518 | 0.2281 | 0.2438 | 0.2680 |
| | LCL | 0.1133 | 0.1870 | 0.1878 | 0.1687 | | LCL | 0.0450 | 0.1719 | 0.1562 | 0.1320 |

**Table 5:** *ARL* performance for the two-sided Shewhart and EWMA charts, Simplex model.

| $\sigma_0$ | $\mu_1$ | SH | $\lambda=0.05$ | $\lambda=0.10$ | $\lambda=0.20$ | $\sigma_0$ | $\mu_1$ | SH | $\lambda=0.05$ | $\lambda=0.10$ | $\lambda=0.20$ |
|---|---|---|---|---|---|---|---|---|---|---|---|
| 0.37 | 0.12 | **1.09** | 4.01 | 3.34 | 3.03 | 0.71 | 0.12 | 5.94 | **3.03** | 5.61 | 5.05 |
| | 0.14 | **2.15** | 4.86 | 4.20 | 3.65 | | 0.14 | 19.22 | **3.65** | 7.54 | 7.22 |
| | 0.16 | 9.34 | 6.84 | 5.95 | **5.38** | | 0.16 | 64.92 | **5.38** | 12.46 | 14.54 |
| | 0.18 | 70.24 | 14.04 | **13.34** | 15.08 | | 0.18 | 213.70 | **15.08** | 40.03 | 78.29 |
| | 0.20 | 370.46 | 371.48 | 366.76 | 373.87 | | 0.20 | 370.35 | 373.87 | 371.47 | 371.57 |
| | 0.22 | 55.04 | 14.15 | **13.19** | 13.41 | | 0.22 | 135.09 | **13.41** | 31.77 | 35.78 |
| | 0.24 | 10.52 | 7.01 | 6.08 | **5.48** | | 0.24 | 42.76 | **5.48** | 12.38 | 12.12 |
| | 0.26 | **3.71** | 4.98 | 4.28 | 3.80 | | 0.26 | 17.56 | **3.80** | 7.73 | 7.21 |
| | 0.28 | **1.98** | 4.02 | 3.52 | 3.11 | | 0.28 | 8.93 | **3.11** | 5.83 | 5.27 |
| | UCL | 0.2784 | 0.2094 | 0.2146 | 0.2225 | | UCL | 0.3586 | 0.2178 | 0.2277 | 0.2429 |
| | LCL | 0.1379 | 0.1909 | 0.1854 | 0.1775 | | LCL | 0.0969 | 0.1822 | 0.1723 | 0.1571 |
| $\sigma_0$ | $\mu_1$ | SH | $\lambda=0.05$ | $\lambda=0.10$ | $\lambda=0.20$ | $\sigma_0$ | $\mu_1$ | SH | $\lambda=0.05$ | $\lambda=0.10$ | $\lambda=0.20$ |
| 0.50 | 0.12 | **1.87** | 3.34 | 4.21 | 3.66 | 1.20 | 0.12 | 35.02 | **4.90** | 9.63 | 10.69 |
| | 0.14 | 5.72 | **4.20** | 5.37 | 4.76 | | 0.14 | 80.59 | **6.17** | 14.46 | 21.26 |
| | 0.16 | 24.83 | **5.95** | 8.16 | 7.83 | | 0.16 | 173.86 | **9.08** | 29.74 | 74.03 |
| | 0.18 | 128.23 | **13.34** | 21.06 | 28.78 | | 0.18 | 332.35 | **20.24** | 125.34 | 546.56 |
| | 0.20 | 370.48 | 366.76 | 375.08 | 371.43 | | 0.20 | 370.40 | 373.08 | 370.88 | 370.23 |
| | 0.22 | 90.00 | **13.19** | 19.51 | 21.54 | | 0.22 | 191.01 | **20.17** | 60.46 | 67.19 |
| | 0.24 | 21.68 | **6.08** | 8.26 | 7.72 | | 0.24 | 84.09 | **9.25** | 23.01 | 24.83 |
| | 0.26 | 7.94 | **4.28** | 5.52 | 4.94 | | 0.26 | 41.59 | **6.36** | 13.65 | 13.68 |
| | 0.28 | 3.94 | **3.52** | 4.38 | 3.83 | | 0.28 | 23.42 | **5.01** | 9.75 | 9.26 |
| | UCL | 0.3085 | 0.2126 | 0.2197 | 0.2304 | | UCL | 0.4743 | 0.2296 | 0.2461 | 0.2725 |
| | LCL | 0.1205 | 0.1874 | 0.1803 | 0.1696 | | LCL | 0.0594 | 0.1704 | 0.1539 | 0.1275 |

**Table 6:** *ARL* performance for the two-sided Shewhart and EWMA charts, Unit Gamma model.

| $\tau_0$ | $\mu_1$ | SH | $\lambda=0.05$ | $\lambda=0.10$ | $\lambda=0.20$ | $\tau_0$ | $\mu_1$ | SH | $\lambda=0.05$ | $\lambda=0.10$ | $\lambda=0.20$ |
|---|---|---|---|---|---|---|---|---|---|---|---|
| 155 | 0.12 | **1.40** | 4.23 | 3.65 | 3.21 | 51 | 0.12 | **4.85** | 6.56 | 5.67 | 5.10 |
|  | 0.14 | **2.85** | 5.26 | 4.53 | 3.96 |  | 0.14 | 12.31 | 8.60 | 7.57 | **7.21** |
|  | 0.16 | 10.00 | 7.49 | 6.54 | **6.01** |  | 0.16 | 38.78 | 13.27 | **12.41** | 13.56 |
|  | 0.18 | 63.21 | 15.74 | **15.09** | 16.94 |  | 0.18 | 144.77 | **32.33** | 37.36 | 58.06 |
|  | 0.20 | 370.30 | 370.39 | 370.58 | 370.54 |  | 0.20 | 370.37 | 370.22 | 370.64 | 370.62 |
|  | 0.22 | 90.28 | 15.91 | **15.10** | 16.01 |  | 0.22 | 213.14 | **32.10** | 34.10 | 41.43 |
|  | 0.24 | 17.82 | 7.60 | 6.63 | **6.02** |  | 0.24 | 78.28 | 13.46 | **12.48** | 12.90 |
|  | 0.26 | 5.36 | 5.29 | 4.60 | **4.06** |  | 0.26 | 31.52 | 8.75 | 7.73 | **7.28** |
|  | 0.28 | **2.36** | 4.26 | 3.71 | 3.26 |  | 0.28 | 14.37 | 6.64 | 5.80 | **5.24** |
|  | UCL | 0.2882 | 0.2103 | 0.2160 | 0.2247 |  | UCL | 0.3662 | 0.2179 | 0.2278 | 0.2431 |
|  | LCL | 0.1330 | 0.1897 | 0.1840 | 0.1753 |  | LCL | 0.0961 | 0.1821 | 0.1722 | 0.1569 |
| $\tau_0$ | $\mu_1$ | SH | $\lambda=0.05$ | $\lambda=0.10$ | $\lambda=0.20$ | $\tau_0$ | $\mu_1$ | SH | $\lambda=0.05$ | $\lambda=0.10$ | $\lambda=0.20$ |
| 96 | 0.12 | **2.23** | 5.05 | 4.35 | 3.79 | 20 | 0.12 | 14.82 | 10.22 | **9.22** | 9.47 |
|  | 0.14 | 5.36 | 6.43 | 5.58 | **4.99** |  | 0.14 | 34.21 | 14.10 | **13.56** | 15.87 |
|  | 0.16 | 18.84 | 9.44 | 8.39 | **8.22** |  | 0.16 | 85.15 | **23.67** | 25.25 | 40.62 |
|  | 0.18 | 95.63 | **20.95** | 21.59 | 27.55 |  | 0.18 | 210.67 | **65.20** | 92.48 | 219.01 |
|  | 0.20 | 370.37 | 370.50 | 370.30 | 370.97 |  | 0.20 | 370.37 | 370.05 | 370.49 | 370.41 |
|  | 0.22 | 138.59 | 21.44 | **20.94** | 24.58 |  | 0.22 | 316.76 | **60.87** | 67.42 | 79.77 |
|  | 0.24 | 35.74 | 9.54 | 8.52 | **8.21** |  | 0.24 | 184.20 | 23.67 | **23.59** | 27.75 |
|  | 0.26 | 11.80 | 6.50 | 5.64 | **5.11** |  | 0.26 | 102.06 | 14.31 | **13.35** | 14.13 |
|  | 0.28 | 4.97 | 5.12 | 4.42 | **3.92** |  | 0.28 | 58.31 | 10.46 | 9.43 | **9.34** |
|  | UCL | 0.3155 | 0.2131 | 0.2203 | 0.2430 |  | UCL | 0.4878 | 0.2284 | 0.2443 | 0.2690 |
|  | LCL | 0.1184 | 0.1869 | 0.1797 | 0.1570 |  | LCL | 0.0596 | 0.1716 | 0.1557 | 0.1310 |

From Table 4 we notice that the use of the two-sided EWMA offers an improved detection ability (compared to the Shewhart chart) of changes in process mean level. Except for large shifts in $\mu_0$, and only in Case 1, the EWMA chart with $\lambda = 0.05$ is the best chart for detecting shifts $\mu_1 \in [0.18, 0.22]$ (Case 2, 3, and 4) and as the $\mu_1$ value increases, a larger $\lambda$, like 0.10 or 0.20, gives an improved performance. Similar conclusions are reached from Table 6 and the performance of the two-sided EWMA for the Unit Gamma distribution.

Note also that the EWMA chart for the Simplex distribution (Table 5) attains the minimum $ARL_1$ in almost all cases for $\lambda = 0.05$ except for the charts in Case 1, where for small shifts the EWMA with $\lambda = 0.10$ has the best performance; for moderate shifts, the EWMA chart with value $\lambda = 0.20$ is suggested. For large increasing or decreasing shifts in Case 1 scenario, the Shewhart chart based on the Simplex distribution has the best performance.

Next, we present the results of a robustness study, where the performance of each EWMA chart, for a given value of $\lambda$, is examined when the limits of a "wrong model" are used, while the observations are from the "true model". Following the work of Ho et al. (2019), the performance of the EWMA chart is examined in terms of $ARL$, $SDRL$ and $MRL$. The aim is to investigate how much the EWMA chart is affected by using "wrong limits", i.e. limits that have been obtained under a different model. As already said, according to Table 3 and Figure 1, for each of the Cases 1-4, the different models cannot be considered as very different between each other.

The results of a simulation study (based on 10000 simulation runs) is presented in the following Tables 7-15. In each of these tables we provide the $ARL$, $SDRL$ and $MRL$ values under the 'correct' (or 'true') probability model. These are labelled as 'Beta True' (Tables 7-9), 'Simplex True' (Tables 10-12) and 'Unit Gamma True' (Tables 13-15). The remaining six columns consist of the values for the $ARL$, $SDRL$ and $MRL$ when the control limits have been evaluated under the 'wrong' probability model.

From Tables 7-15 we notice that mainly, the IC performance of the two-sided EWMA charts is affected by the use of wrong control limits. Significant differences are also present in the OOC ARL, mainly for shifts in the interval [0.18, 0.22].

Regarding the IC $ARL$, its value is either below or above the nominal $ARL_0$ value of 370.4 when 'wrong' limits are used. It is known that a large IC $ARL$ value is, in general, desirable unless the OOC performance is affected. For example, a larger (than the desired) IC $ARL$ might result in larger OOC $ARL$ values, which means that there are delays in the time needed for detecting a shift in $\mu_0$. On the other hand, an IC $ARL$ value lower than the nominal $ARL_0$ value results in an excessive number of false alarms and frequent, unnecessary, stops of process monitoring. For example, except for Case 2, the EWMA chart with Simplex limits has also a much different IC performance when the true model is that of the Beta distribution.

In terms of the OOC performance, we notice that the use of wrong limits leads to deterioration of the chart's power (i.e., increase in the $ARL_1$ values) especially when the limits that are based on the Unit Gamma model are used when the true process is a Beta one (see Tables 7-9). In Cases 1 and 3 we see the most noticeable deviations. As $\lambda$ increases, the difference between the actual and the desired IC $ARL$ value becomes greater. See, for example, Table 9.

On the other hand, the $ARL_1$ values are lower than the ones under the true model, when the control limits of the Simplex model are used, mainly in Case 2. Note also, that this is because the IC $ARL$ value is lower than the nominal value of 370.4. Therefore, an increase in false alarm rates results in a decrease in the OOC $ARL$ values. The use of the limits obtained under the Simplex models results in a significant difference in IC and OOC ARL values (in the interval [0.18, 22]) in all cases, except for Case 2. For the Simplex two-sided EWMA chart and for $\lambda = 0.05$ (Table 10) we notice lower $ARL_1$ values when we use the control limits under the Beta model. However, the IC $ARL$ of the chart is lower than the nominal value.

From Tables 11 and 12 (i.e., for $\lambda = 0.10$ and $0.20$, respectively), we notice that the $ARL_0$ is larger for the Beta and Unit Gamma models (the 'wrong' ones). However, the $ARL_1$ values do not differ significantly than the one under the Simplex model (the 'right' one), except for Case 4. In Table 12, we also notice that for $\mu_1 = 0.18$, the $ARL_1 > ARL_0$ and thus, the two-sided EWMA is $ARL$-biased. This is attributed to the skewness of the $S(0.2,1.2^2)$ distribution, which is the largest one, compared to all other cases (see Table 3). A control chart is said to have an $ARL$-biased performance when it takes longer (in average) to detect some shifts in the IC value of the monitored parameter than give a false alarm (Pignatiello et al. (1995), Knoth and Morais (2015)). A two-sided control chart has $ARL$-biased performance when the distribution of the statistic plotted on the chart is not symmetric and when symmetric limits are used. It is worth mentioning that our numerical analysis showed that as the value of $\lambda$ increases and for distributional models that deviate significantly from symmetry, the $ARL$ performance of the two-sided EWMA chart is biased. However, for $\lambda = 0.05$ or $0.10$, the bias in the $ARL$ reduces. For the considered shifts in $\mu_0$, only in the results for Case 4, in Table 12, we notice an $ARL$-biased performance.

A possible solution to the problem of $ARL$-biased performance would be to consider asymmetric limits, i.e., instead of using the same value $L$ in the formulas of the control limits of the two-sided EWMA chart use $L_U, L_L$ (with $L_U \neq L_L$) for the $UCL$ and the $LCL$, respectively. However, the aim of this work is not to investigate under which circumstances the two-sided EWMA charts are $ARL$-biased as well as how to deal with this problem. This will be done separately in a future paper.

Finally, for the two-sided EWMA chart under the Unit Gamma model we notice that, especially for Cases 1 and 2, there is a significant deviation mainly in the IC performance of the chart, when the limits of the Simplex or the Beta distribution ('wrong' model) are used.

From Tables 13 and 14 we notice that the use of control limits under the Simplex model results in much more different $ARL$ values, especially in Cases 1 and 4.

In some cases, for moderate to large shifts in $\mu_0$ we do not notice significant deviation between the 'correct' and the 'wrong' model. See for example Table 7, Case 1 where the wrong model is the Simplex distribution. Or, Table 10, Case 4, when the wrong model is either the Beta or the Unit Gamma distribution as well as Table 14, Case 3 when the wrong models is either the Beta or the Simplex distribution. Therefore, for small shifts in process mean level and in the IC case, the performance of the two-sided EWMA chart is affected by using wrong limits (i.e., limits that have been obtained under a similar but different model) whereas, for large shifts, this effect declines and almost no difference is noticed. Moreover, the differences between the three models are not significant in Case 3. In Case 4 we have the largest variance among the considered four Cases and the differences are significant in both IC and OOC cases (small shifts), especially for $\lambda = 0.05$. Case 1 has the smallest variance and in almost all the examined scenarios, only the IC performance is the one that it is affected by the use of wrong limits. This has been also noticed by Ho et al. (2019) in the case of Shewhart-type charts.

It is worth mentioning that the limits for each chart are given with four decimals accuracy. Even though they are very close, there are small differences which result in the differences in the performance of the charts, especially for an IC process. Clearly, the determination of the appropriate model is important. We would advise practitioners to apply the EWMA chart with $\lambda = 0.05$ since it has better performance than the Shewhart chart in small increasing and decreasing shifts, while its OOC performance can be considered to be the same, whether the correct limits are used or not. However, the price to pay for this simple solution to this problem (i.e., without first select the model with the best fit) is that the IC performance can be much different than the desired one.

**Table 7:** Performance of two-sided EWMA Charts, $\lambda = 0.05$, True Model: Beta distribution.

| | | Beta - True | | | Simplex | | | Unit Gamma | | |
|---|---|---|---|---|---|---|---|---|---|---|
| | $\mu$ | ARL | SDRL | MRL | ARL | SDRL | MRL | ARL | SDRL | MRL |
| Case 1: | 0.12 | 3.98 | 0.47 | 4 | 3.88 | 0.50 | 4 | 4.15 | 0.46 | 4 |
| $\phi_0 = 290$ | 0.14 | 4.87 | 0.75 | 5 | 4.75 | 0.74 | 5 | 5.13 | 0.78 | 5 |
| $\sigma_0 = 0.37$ | 0.16 | 6.81 | 1.45 | 7 | 6.63 | 1.44 | 6 | 7.26 | 1.56 | 7 |
| $\tau_0 = 155$ | 0.18 | 13.89 | 5.19 | 13 | 13.44 | 5.06 | 12 | 15.19 | 5.56 | 14 |
| | 0.20 | 370.14 | 357.48 | 257 | 302.44 | 287.78 | 216 | 564.09 | 546.54 | 395 |
| | 0.22 | 14.04 | 5.63 | 13 | 13.49 | 5.36 | 12 | 15.36 | 6.16 | 14 |
| | 0.24 | 6.87 | 1.68 | 7 | 6.65 | 1.65 | 6 | 7.34 | 1.79 | 7 |
| | 0.26 | 4.90 | 0.91 | 5 | 4.77 | 0.89 | 5 | 5.15 | 0.96 | 5 |
| | 0.28 | 3.97 | 0.63 | 4 | 3.89 | 0.63 | 4 | 4.18 | 0.65 | 4 |
| | LCL | 0.1907 | | | 0.1906 | | | 0.1897 | | |
| | UCL | 0.2093 | | | 0.2094 | | | 0.2103 | | |
| Case 2: | 0.12 | 5.04 | 0.77 | 5 | 5.01 | 0.76 | 5 | 5.02 | 0.77 | 5 |
| $\phi = 148$ | 0.14 | 6.42 | 1.23 | 6 | 6.40 | 1.24 | 6 | 6.38 | 1.27 | 6 |
| $\sigma_0 = 0.50$ | 0.16 | 9.46 | 2.58 | 9 | 9.41 | 2.58 | 9 | 9.40 | 2.55 | 9 |
| $\tau_0 = 96$ | 0.18 | 21.09 | 9.89 | 19 | 20.95 | 10.01 | 19 | 20.97 | 9.84 | 19 |
| | 0.20 | 370.14 | 351.44 | 265 | 363.33 | 343.52 | 259 | 357.90 | 345.13 | 253 |
| | 0.22 | 21.05 | 10.73 | 19 | 21 | 10.59 | 19 | 21.13 | 10.74 | 19 |
| | 0.24 | 9.55 | 3.02 | 9 | 9.51 | 2.98 | 9 | 9.49 | 2.97 | 9 |
| | 0.26 | 6.47 | 1.56 | 6 | 6.45 | 1.55 | 6 | 6.46 | 1.56 | 6 |
| | 0.28 | 5.09 | 1.02 | 5 | 5.08 | 1 | 5 | 5.10 | 1.01 | 5 |
| | LCL | 0.1870 | | | 0.1874 | | | 0.1869 | | |
| | UCL | 0.2130 | | | 0.2126 | | | 0.2131 | | |
| Case 3: | 0.12 | 6.49 | 1.20 | 6 | 6.59 | 1.23 | 6 | 6.59 | 1.22 | 6 |
| $\phi = 80$ | 0.14 | 8.54 | 2.10 | 8 | 8.68 | 2.11 | 8 | 8.62 | 2.08 | 8 |
| $\sigma_0 = 0.71$ | 0.16 | 13.13 | 4.59 | 12 | 13.40 | 4.61 | 13 | 13.43 | 4.57 | 13 |
| $\tau_0 = 51$ | 0.18 | 31.81 | 18.26 | 27 | 33.07 | 18.86 | 28 | 32.73 | 19.02 | 28 |
| | 0.20 | 370.48 | 357.90 | 261 | 410.39 | 394.24 | 289 | 410.51 | 394.74 | 289 |
| | 0.22 | 31.70 | 19.05 | 27 | 32.48 | 20.19 | 27 | 32.42 | 19.80 | 27 |
| | 0.24 | 13.28 | 5.28 | 12 | 13.48 | 5.24 | 12 | 13.51 | 5.29 | 13 |
| | 0.26 | 8.65 | 2.59 | 8 | 8.80 | 2.67 | 8 | 8.79 | 2.63 | 8 |
| | 0.28 | 6.58 | 1.63 | 6 | 6.69 | 1.65 | 6 | 6.67 | 1.67 | 6 |
| | LCL | 0.1823 | | | 0.1822 | | | 0.1821 | | |
| | UCL | 0.2177 | | | 0.2178 | | | 0.2179 | | |
| Case 4: | 0.12 | 10.15 | 2.69 | 10 | 10.45 | 2.67 | 10 | 10.07 | 2.61 | 10 |
| $\phi = 31$ | 0.14 | 13.93 | 4.71 | 13 | 14.52 | 4.96 | 14 | 13.85 | 4.66 | 13 |
| $\sigma_0 = 1.20$ | 0.16 | 23.24 | 11.01 | 21 | 24.34 | 11.64 | 22 | 23.09 | 11.03 | 21 |
| $\tau_0 = 20$ | 0.18 | 64.30 | 46.81 | 52 | 69.74 | 52.14 | 55 | 63.25 | 46.57 | 50 |
| | 0.20 | 370.50 | 360.16 | 257 | 439.72 | 424.28 | 312 | 351.96 | 344.37 | 243 |
| | 0.22 | 59.48 | 44.65 | 47 | 64.97 | 50.17 | 50.5 | 59.27 | 44.36 | 47 |
| | 0.24 | 23.13 | 12.74 | 20 | 24.21 | 13.02 | 21 | 23.15 | 12.42 | 20 |
| | 0.26 | 14.17 | 5.92 | 13 | 14.65 | 6.07 | 13 | 14.04 | 5.92 | 13 |
| | 0.28 | 10.37 | 3.62 | 10 | 10.69 | 3.71 | 10 | 10.27 | 3.61 | 10 |
| | LCL | 0.1719 | | | 0.1704 | | | 0.1716 | | |
| | UCL | 0.2281 | | | 0.2296 | | | 0.2284 | | |

**Table 8:** Performance of two-sided EWMA Charts, $\lambda = 0.10$, True Model: Beta distribution.

|  | $\mu$ | Beta - True | | | Simplex | | | Unit Gamma | | |
|---|---|---|---|---|---|---|---|---|---|---|
|  |  | ARL | SDRL | MRL | ARL | SDRL | MRL | ARL | SDRL | MRL |
| Case 1: | 0.12 | 3.40 | 0.51 | 3 | 3.48 | 0.53 | 3 | 3.65 | 0.54 | 4 |
| $\phi_0 = 290$ | 0.14 | 4.23 | 0.73 | 4 | 4.32 | 0.73 | 4 | 4.53 | 0.77 | 4 |
| $\sigma_0 = 0.37$ | 0.16 | 5.96 | 1.47 | 6 | 6.13 | 1.52 | 6 | 6.51 | 1.62 | 6 |
| $\tau_0 = 155$ | 0.18 | 13.20 | 6.08 | 12 | 13.91 | 6.44 | 12 | 15.34 | 7.16 | 14 |
|  | 0.20 | 370.30 | 359.15 | 262 | 462.27 | 454.45 | 322 | 766.51 | 741.72 | 538 |
|  | 0.22 | 13.29 | 6.56 | 12 | 13.87 | 6.78 | 12 | 15.29 | 7.64 | 14 |
|  | 0.24 | 6.06 | 1.72 | 6 | 6.18 | 1.71 | 6 | 6.59 | 1.86 | 6 |
|  | 0.26 | 4.25 | 0.88 | 4 | 4.36 | 0.91 | 4 | 4.57 | 0.95 | 4 |
|  | 0.28 | 3.46 | 0.60 | 3 | 3.53 | 0.61 | 3 | 3.68 | 0.63 | 4 |
|  | $LCL$ | 0.1855 | | | 0.1854 | | | 0.1840 | | |
|  | $UCL$ | 0.2145 | | | 0.2146 | | | 0.2160 | | |
| Case 2: | 0.12 | 4.35 | 0.72 | 4 | 4.31 | 0.72 | 4 | 4.32 | 0.72 | 4 |
| $\phi = 148$ | 0.14 | 5.52 | 1.21 | 5 | 5.48 | 1.21 | 5 | 5.50 | 1.23 | 5 |
| $\sigma_0 = 0.50$ | 0.16 | 8.40 | 2.73 | 8 | 8.30 | 2.74 | 8 | 8.25 | 2.67 | 8 |
| $\tau_0 = 96$ | 0.18 | 21.40 | 12.77 | 18 | 20.96 | 12.29 | 18 | 20.92 | 12.28 | 18 |
|  | 0.20 | 370.44 | 362.60 | 257 | 345.56 | 336.98 | 244 | 331.58 | 323 | 233 |
|  | 0.22 | 21.19 | 13.21 | 18 | 20.60 | 12.50 | 17 | 20.55 | 12.83 | 17 |
|  | 0.24 | 8.47 | 3.16 | 8 | 8.41 | 3.19 | 8 | 8.46 | 3.22 | 8 |
|  | 0.26 | 5.64 | 1.52 | 5 | 5.59 | 1.54 | 5 | 5.56 | 1.52 | 5 |
|  | 0.28 | 4.41 | 0.98 | 4 | 4.37 | 0.97 | 4 | 4.38 | 0.99 | 4 |
|  | $LCL$ | 0.1898 | | | 0.1803 | | | 0.1797 | | |
|  | $UCL$ | 0.2202 | | | 0.2197 | | | 0.2203 | | |
| Case 3: | 0.12 | 5.62 | 1.18 | 5 | 5.73 | 1.23 | 6 | 5.71 | 1.21 | 6 |
| $\phi = 80$ | 0.14 | 7.55 | 2.16 | 7 | 7.65 | 2.17 | 7 | 7.70 | 2.20 | 7 |
| $\sigma_0 = 0.71$ | 0.16 | 12.15 | 5.13 | 11 | 12.62 | 5.33 | 11 | 12.49 | 5.24 | 11 |
| $\tau_0 = 51$ | 0.18 | 35.86 | 25.51 | 29 | 37.49 | 26.48 | 30 | 38.10 | 27.78 | 30 |
|  | 0.20 | 370.49 | 363.44 | 258 | 410.86 | 398.89 | 289 | 416.80 | 405.05 | 294 |
|  | 0.22 | 33.62 | 24.82 | 27 | 34.74 | 25.64 | 27 | 35.23 | 25.55 | 28 |
|  | 0.24 | 12.15 | 5.78 | 11 | 12.50 | 5.96 | 11 | 12.71 | 6.15 | 11 |
|  | 0.26 | 7.62 | 2.71 | 7 | 7.77 | 2.75 | 7 | 7.77 | 2.74 | 7 |
|  | 0.28 | 5.77 | 1.63 | 6 | 5.81 | 1.65 | 6 | 5.82 | 1.65 | 6 |
|  | $LCL$ | 0.1725 | | | 0.1723 | | | 0.1722 | | |
|  | $UCL$ | 0.2275 | | | 0.2277 | | | 0.2278 | | |
| Case 4: | 0.12 | 9.11 | 2.82 | 9 | 9.70 | 3.06 | 9 | 9.21 | 2.88 | 9 |
| $\phi = 31$ | 0.14 | 13.28 | 5.52 | 12 | 14.40 | 6.17 | 13 | 13.34 | 5.67 | 12 |
| $\sigma_0 = 1.20$ | 0.16 | 24.96 | 15.17 | 21 | 27.69 | 16.92 | 23 | 25.05 | 14.98 | 21 |
| $\tau_0 = 20$ | 0.18 | 87.98 | 76.03 | 65 | 112.78 | 98.01 | 83 | 89.84 | 76.45 | 67 |
|  | 0.20 | 370.66 | 359.86 | 262 | 518.58 | 507.70 | 365 | 380.12 | 372.38 | 268 |
|  | 0.22 | 66.36 | 57.38 | 48 | 77.23 | 67.79 | 57 | 68.35 | 59.22 | 50 |
|  | 0.24 | 23.07 | 15.16 | 19 | 25.87 | 17.30 | 21 | 23.54 | 15.68 | 19 |
|  | 0.26 | 13.30 | 6.73 | 12 | 14.13 | 7.22 | 12 | 13.25 | 6.75 | 12 |
|  | 0.28 | 9.30 | 3.86 | 9 | 9.93 | 4.18 | 9 | 9.40 | 3.90 | 9 |
|  | $LCL$ | 0.1562 | | | 0.1539 | | | 0.1557 | | |
|  | $UCL$ | 0.2438 | | | 0.2461 | | | 0.2443 | | |

**Table 9:** Performance of two-sided EWMA Charts, $\lambda = 0.20$, True Model: Beta distribution.

| | | Beta - True | | | Simplex | | | Unit Gamma | | |
|---|---|---|---|---|---|---|---|---|---|---|
| | $\mu$ | ARL | SDRL | MRL | ARL | SDRL | MRL | ARL | SDRL | MRL |
| Case 1: | 0.12 | 3.06 | 0.38 | 3 | 3.04 | 0.38 | 3 | 3.22 | 0.43 | 3 |
| $\phi_0 = 290$ | 0.14 | 3.67 | 0.73 | 4 | 3.64 | 0.72 | 4 | 4.01 | 0.81 | 4 |
| $\sigma_0 = 0.37$ | 0.16 | 5.32 | 1.62 | 5 | 5.25 | 1.59 | 5 | 6.06 | 1.87 | 6 |
| $\tau_0 = 155$ | 0.18 | 14.10 | 8.66 | 12 | 13.78 | 8.31 | 12 | 19.26 | 12.89 | 16 |
| | 0.20 | 370.88 | 368.01 | 256 | 329.78 | 323.59 | 234 | 971.99 | 920.30 | 685 |
| | 0.22 | 13.89 | 8.67 | 12 | 13.33 | 8.25 | 11 | 17.85 | 11.93 | 15 |
| | 0.24 | 5.42 | 1.86 | 5 | 5.35 | 1.84 | 5 | 6.14 | 2.19 | 6 |
| | 0.26 | 3.73 | 0.88 | 4 | 3.68 | 0.86 | 4 | 4.08 | 0.96 | 4 |
| | 0.28 | 3.06 | 0.57 | 3 | 3.04 | 0.55 | 3 | 3.26 | 0.59 | 3 |
| | LCL | 0.1776 | | | 0.1775 | | | 0.1753 | | |
| | UCL | 0.2224 | | | 0.2225 | | | 0.2247 | | |
| Case 2: | 0.12 | 3.81 | 0.76 | 4 | 3.68 | 0.73 | 4 | 3.77 | 0.75 | 4 |
| $\phi = 148$ | 0.14 | 4.94 | 1.34 | 5 | 4.78 | 1.28 | 5 | 4.92 | 1.34 | 5 |
| $\sigma_0 = 0.50$ | 0.16 | 8.11 | 3.41 | 7 | 7.55 | 3.13 | 7 | 7.97 | 3.40 | 7 |
| $\tau_0 = 96$ | 0.18 | 27.07 | 20.43 | 21 | 23.94 | 17.42 | 19 | 26.64 | 20.07 | 21 |
| | 0.20 | 370.32 | 366.88 | 261 | 269.88 | 263.70 | 186 | 346.86 | 340.11 | 242 |
| | 0.22 | 24.19 | 18.83 | 19 | 21.60 | 16.23 | 17 | 23.87 | 18.27 | 18 |
| | 0.24 | 8.12 | 3.89 | 7 | 7.70 | 3.61 | 7 | 8 | 3.82 | 7 |
| | 0.26 | 5.09 | 1.71 | 5 | 4.86 | 1.56 | 5 | 5.02 | 1.68 | 5 |
| | 0.28 | 3.88 | 0.97 | 4 | 3.77 | 0.94 | 4 | 3.87 | 0.97 | 4 |
| | LCL | 0.1687 | | | 0.1696 | | | 0.1570 | | |
| | UCL | 0.2313 | | | 0.2304 | | | 0.2430 | | |
| Case 3: | 0.12 | 5.05 | 1.31 | 5 | 5.11 | 1.34 | 5 | 5.11 | 1.34 | 5 |
| $\phi = 80$ | 0.14 | 7.04 | 2.57 | 6 | 7.18 | 2.64 | 7 | 7.24 | 2.69 | 7 |
| $\sigma_0 = 0.71$ | 0.16 | 13.19 | 7.43 | 11 | 13.58 | 7.90 | 11 | 13.54 | 7.67 | 12 |
| $\tau_0 = 51$ | 0.18 | 55.18 | 47.41 | 41 | 58.76 | 51.56 | 43 | 58.47 | 51.48 | 43 |
| | 0.20 | 370.79 | 366.72 | 256 | 411.77 | 406.29 | 287 | 406.40 | 409.79 | 281 |
| | 0.22 | 40.80 | 34.67 | 31 | 42.51 | 36.37 | 31 | 42.27 | 36.32 | 32 |
| | 0.24 | 12.46 | 7.60 | 10 | 12.93 | 8.19 | 11 | 12.91 | 7.93 | 11 |
| | 0.26 | 7.20 | 3.20 | 6 | 7.24 | 3.16 | 7 | 7.24 | 3.23 | 7 |
| | 0.28 | 5.16 | 1.75 | 5 | 5.20 | 1.76 | 5 | 5.20 | 1.78 | 5 |
| | LCL | 0.1575 | | | 0.1571 | | | 0.1569 | | |
| | UCL | 0.2425 | | | 0.2429 | | | 0.2431 | | |
| Case 4: | 0.12 | 9.21 | 3.91 | 8 | 10.02 | 4.33 | 9 | 9.44 | 4.01 | 8 |
| $\phi = 31$ | 0.14 | 15.30 | 8.97 | 13 | 17.18 | 10.39 | 14 | 15.96 | 9.48 | 13 |
| $\sigma_0 = 1.20$ | 0.16 | 37.61 | 30.51 | 28 | 47.64 | 38.74 | 36 | 41.05 | 33 | 31 |
| $\tau_0 = 20$ | 0.18 | 188.96 | 181.58 | 133 | 281.96 | 275.28 | 196 | 216.43 | 205.76 | 154 |
| | 0.20 | 370.16 | 364.37 | 258 | 505.41 | 498.89 | 351.5 | 415.54 | 410.62 | 292 |
| | 0.22 | 78.97 | 72.17 | 58 | 96.57 | 90.87 | 69 | 85.65 | 81.30 | 60 |
| | 0.24 | 26.48 | 21.16 | 20 | 31.30 | 25.70 | 24 | 27.90 | 22.24 | 21 |
| | 0.26 | 13.57 | 8.70 | 11 | 14.96 | 9.92 | 12 | 14.12 | 9.30 | 11 |
| | 0.28 | 9.02 | 4.76 | 8 | 9.64 | 5.09 | 8 | 9.23 | 4.85 | 8 |
| | LCL | 0.1320 | | | 0.1275 | | | 0.1310 | | |
| | UCL | 0.2680 | | | 0.2725 | | | 0.2690 | | |

**Table 10:** Performance of two-sided EWMA Charts, $\lambda = 0.05$, True Model: Simplex distribution.

| | | Simplex - True | | | Beta | | | Unit Gamma | | |
|---|---|---|---|---|---|---|---|---|---|---|
| | $\mu$ | ARL | SDRL | MRL | ARL | SDRL | MRL | ARL | SDRL | MRL |
| Case 1: | 0.12 | 4.01 | 0.25 | 4 | 3.93 | 0.30 | 4 | 4.09 | 0.30 | 4 |
| $\phi_0 = 290$ | 0.14 | 4.86 | 0.60 | 5 | 4.70 | 0.61 | 5 | 5.10 | 0.61 | 5 |
| $\sigma_0 = 0.37$ | 0.16 | 6.84 | 1.25 | 7 | 6.58 | 1.23 | 6 | 7.23 | 1.30 | 7 |
| $\tau_0 = 155$ | 0.18 | 14.04 | 4.87 | 13 | 13.39 | 4.76 | 13 | 15.12 | 5.30 | 14 |
| | 0.20 | 371.48 | 359.17 | 260 | 292.12 | 277.86 | 206 | 548.49 | 523.70 | 391 |
| | 0.22 | 14.15 | 6.01 | 13 | 13.44 | 5.73 | 12 | 15.28 | 6.42 | 14 |
| | 0.24 | 7.01 | 1.94 | 7 | 6.74 | 1.88 | 6 | 7.41 | 2.02 | 7 |
| | 0.26 | 4.98 | 1.11 | 5 | 4.83 | 1.07 | 5 | 5.20 | 1.14 | 5 |
| | 0.28 | 4.02 | 0.77 | 4 | 3.93 | 0.76 | 4 | 4.19 | 0.79 | 4 |
| | LCL | 0.1906 | | | 0.1907 | | | 0.1897 | | |
| | UCL | 0.2094 | | | 0.2093 | | | 0.2103 | | |
| Case 2: | 0.12 | 3.34 | 0.53 | 5 | 3.45 | 0.52 | 5 | 3.67 | 0.51 | 5 |
| $\phi = 148$ | 0.14 | 4.20 | 0.90 | 6 | 4.28 | 0.92 | 6 | 4.49 | 0.93 | 6 |
| $\sigma_0 = 0.50$ | 0.16 | 5.95 | 2.10 | 9 | 6.09 | 2.10 | 9 | 6.50 | 2.15 | 9 |
| $\tau_0 = 96$ | 0.18 | 13.34 | 8.54 | 18 | 13.95 | 9.36 | 19 | 15.42 | 9.30 | 19 |
| | 0.20 | 366.76 | 362.84 | 258 | 449.22 | 421.14 | 311 | 728.90 | 424.44 | 315 |
| | 0.22 | 13.19 | 10.46 | 18 | 13.61 | 10.51 | 19 | 15.01 | 10.84 | 18 |
| | 0.24 | 6.08 | 3.23 | 9 | 6.28 | 3.26 | 9 | 6.65 | 3.27 | 9 |
| | 0.26 | 4.28 | 1.76 | 6 | 4.39 | 1.79 | 6 | 4.61 | 1.78 | 6 |
| | 0.28 | 3.52 | 1.20 | 5 | 3.57 | 1.21 | 5 | 3.73 | 1.20 | 5 |
| | LCL | 0.1874 | | | 0.1870 | | | 0.1869 | | |
| | UCL | 0.2126 | | | 0.2130 | | | 0.2131 | | |
| Case 3: | 0.12 | 3.03 | 0.84 | 6 | 3.02 | 0.83 | 6 | 3.12 | 0.83 | 6 |
| $\phi = 80$ | 0.14 | 3.65 | 1.62 | 8 | 3.58 | 1.60 | 8 | 3.98 | 1.63 | 8 |
| $\sigma_0 = 0.71$ | 0.16 | 5.38 | 3.90 | 12 | 5.25 | 3.96 | 13 | 6.10 | 3.91 | 13 |
| $\tau_0 = 51$ | 0.18 | 15.08 | 18.29 | 28 | 14.06 | 18.31 | 29 | 20.37 | 18.30 | 29 |
| | 0.20 | 373.87 | 356.06 | 263 | 310.78 | 387.26 | 281.5 | 891.57 | 386.01 | 284 |
| | 0.22 | 13.41 | 19.17 | 26 | 12.87 | 20.06 | 27 | 17.00 | 20.24 | 27 |
| | 0.24 | 5.48 | 5.73 | 12 | 5.37 | 6.04 | 12 | 6.11 | 5.86 | 12 |
| | 0.26 | 3.80 | 3.06 | 8 | 3.74 | 3.12 | 8 | 4.14 | 3.16 | 8 |
| | 0.28 | 3.11 | 2.03 | 6 | 3.06 | 2.07 | 6 | 3.31 | 2.07 | 6 |
| | LCL | 0.1822 | | | 0.1823 | | | 0.1821 | | |
| | UCL | 0.2178 | | | 0.2177 | | | 0.2179 | | |
| Case 4: | 0.12 | 4.90 | 1.96 | 10 | 5.01 | 1.88 | 10 | 5.00 | 1.88 | 10 |
| $\phi = 31$ | 0.14 | 6.17 | 4.10 | 14 | 6.33 | 3.90 | 13 | 6.33 | 3.90 | 13 |
| $\sigma_0 = 1.20$ | 0.16 | 9.08 | 10.98 | 22 | 9.32 | 9.95 | 21 | 9.38 | 10.31 | 21 |
| $\tau_0 = 20$ | 0.18 | 20.24 | 54.31 | 59 | 21.30 | 46.96 | 53 | 21.23 | 48.49 | 53 |
| | 0.20 | 373.08 | 360.60 | 266 | 437.96 | 279.63 | 211 | 442.12 | 280.01 | 213 |
| | 0.22 | 20.17 | 45.43 | 46 | 20.84 | 42.64 | 41 | 20.73 | 41.45 | 42 |
| | 0.24 | 9.25 | 14.26 | 20 | 9.52 | 13.64 | 19 | 9.57 | 13.45 | 19 |
| | 0.26 | 6.36 | 7.31 | 13 | 6.53 | 6.95 | 13 | 6.54 | 6.91 | 13 |
| | 0.28 | 5.01 | 4.59 | 10 | 5.13 | 4.49 | 10 | 5.13 | 4.57 | 9 |
| | LCL | 0.1704 | | | 0.1719 | | | 0.1716 | | |
| | UCL | 0.2296 | | | 0.2281 | | | 0.2284 | | |

**Table 11:** Performance of two-sided EWMA Charts, $\lambda = 0.10$, True Model: Simplex distribution.

| | | Simplex - True | | | Beta | | | Unit Gamma | | |
|---|---|---|---|---|---|---|---|---|---|---|
| | $\mu$ | ARL | SDRL | MRL | ARL | SDRL | MRL | ARL | SDRL | MRL |
| Case 1: | 0.12 | 3.34 | 0.48 | 3 | 3.45 | 0.50 | 3 | 3.67 | 0.48 | 4 |
| $\phi_0 = 290$ | 0.14 | 4.20 | 0.55 | 4 | 4.28 | 0.56 | 4 | 4.49 | 0.61 | 4 |
| $\sigma_0 = 0.37$ | 0.16 | 5.95 | 1.26 | 6 | 6.09 | 1.30 | 6 | 6.50 | 1.39 | 6 |
| $\tau_0 = 155$ | 0.18 | 13.34 | 5.85 | 12 | 13.95 | 6.11 | 13 | 15.42 | 6.99 | 14 |
| | 0.20 | 366.76 | 361.67 | 256 | 449.22 | 437.38 | 317 | 728.90 | 694.68 | 512 |
| | 0.22 | 13.19 | 6.73 | 12 | 13.61 | 7.03 | 12 | 15.01 | 7.90 | 13 |
| | 0.24 | 6.08 | 1.95 | 6 | 6.28 | 2.02 | 6 | 6.65 | 2.15 | 6 |
| | 0.26 | 4.28 | 1.05 | 4 | 4.39 | 1.07 | 4 | 4.61 | 1.13 | 4 |
| | 0.28 | 3.52 | 0.72 | 3 | 3.57 | 0.71 | 3 | 3.73 | 0.77 | 4 |
| | LCL | | 0.1854 | | | 0.1855 | | | 0.1840 | |
| | UCL | | 0.2146 | | | 0.2145 | | | 0.2160 | |
| Case 2: | 0.12 | 4.21 | 0.47 | 4 | 4.25 | 0.48 | 4 | 4.26 | 0.48 | 4 |
| $\phi = 148$ | 0.14 | 5.37 | 0.90 | 5 | 5.45 | 0.92 | 5 | 5.45 | 0.93 | 5 |
| $\sigma_0 = 0.50$ | 0.16 | 8.16 | 2.25 | 8 | 8.29 | 2.29 | 8 | 8.30 | 2.29 | 8 |
| $\tau_0 = 96$ | 0.18 | 21.06 | 11.61 | 18 | 21.91 | 12.49 | 19 | 21.92 | 12.26 | 19 |
| | 0.20 | 375.08 | 368.08 | 264 | 416.84 | 410.89 | 288 | 428.18 | 415.18 | 304 |
| | 0.22 | 19.51 | 12.36 | 16 | 20.69 | 12.97 | 17 | 20.20 | 12.66 | 17 |
| | 0.24 | 8.26 | 3.33 | 8 | 8.45 | 3.46 | 8 | 8.42 | 3.37 | 8 |
| | 0.26 | 5.52 | 1.73 | 5 | 5.63 | 1.74 | 5 | 5.64 | 1.77 | 5 |
| | 0.28 | 4.38 | 1.17 | 4 | 4.41 | 1.15 | 4 | 4.43 | 1.15 | 4 |
| | LCL | | 0.1803 | | | 0.1898 | | | 0.1797 | |
| | UCL | | 0.2197 | | | 0.2202 | | | 0.2203 | |
| Case 3: | 0.12 | 5.61 | 0.84 | 6 | 5.66 | 0.84 | 6 | 5.67 | 0.84 | 6 |
| $\phi = 80$ | 0.14 | 7.54 | 1.71 | 7 | 7.61 | 1.72 | 7 | 7.61 | 1.72 | 7 |
| $\sigma_0 = 0.71$ | 0.16 | 12.46 | 4.71 | 11 | 12.69 | 4.75 | 12 | 12.60 | 4.65 | 12 |
| $\tau_0 = 51$ | 0.18 | 40.03 | 28.57 | 32 | 41.65 | 29.74 | 33 | 41.51 | 29.53 | 33 |
| | 0.20 | 371.47 | 361.87 | 261 | 405.57 | 400.67 | 283 | 402.45 | 392.01 | 284 |
| | 0.22 | 31.77 | 23.66 | 25 | 32.34 | 23.90 | 26 | 32.56 | 24.59 | 25 |
| | 0.24 | 12.38 | 6.47 | 11 | 12.47 | 6.52 | 11 | 12.49 | 6.46 | 11 |
| | 0.26 | 7.73 | 3.19 | 7 | 7.89 | 3.23 | 7 | 7.81 | 3.22 | 7 |
| | 0.28 | 5.83 | 2.02 | 5 | 5.88 | 2 | 6 | 5.90 | 2.03 | 6 |
| | LCL | | 0.1723 | | | 0.1725 | | | 0.1722 | |
| | UCL | | 0.2277 | | | 0.2275 | | | 0.2278 | |
| Case 4: | 0.12 | 9.63 | 2.23 | 9 | 9.20 | 2.16 | 9 | 9.23 | 2.19 | 9 |
| $\phi = 31$ | 0.14 | 14.46 | 5.20 | 13 | 13.64 | 4.89 | 13 | 13.69 | 4.94 | 13 |
| $\sigma_0 = 1.20$ | 0.16 | 29.74 | 17.79 | 25 | 26.85 | 15.51 | 23 | 26.94 | 15.84 | 23 |
| $\tau_0 = 20$ | 0.18 | 125.34 | 110.46 | 91 | 104.56 | 88.94 | 77 | 103.40 | 90.40 | 76 |
| | 0.20 | 370.88 | 363.92 | 261 | 310.84 | 305.53 | 216 | 301.09 | 294.40 | 209 |
| | 0.22 | 60.46 | 51.95 | 45 | 54.62 | 47.05 | 40 | 55.43 | 47.55 | 42 |
| | 0.24 | 23.01 | 16.08 | 19 | 21.76 | 15.25 | 18 | 21.95 | 15.25 | 18 |
| | 0.26 | 13.65 | 7.94 | 12 | 12.97 | 7.45 | 11 | 13.05 | 7.60 | 11 |
| | 0.28 | 9.75 | 4.84 | 9 | 9.31 | 4.59 | 8 | 9.31 | 4.66 | 8 |
| | LCL | | 0.1539 | | | 0.1562 | | | 0.1557 | |
| | UCL | | 0.2461 | | | 0.2438 | | | 0.2443 | |

**Table 12:** Performance of two-sided EWMA Charts, $\lambda = 0.20$, True Model: Simplex distribution.

| | | Simplex - True | | | Beta | | | Unit Gamma | | |
|---|---|---|---|---|---|---|---|---|---|---|
| | $\mu$ | ARL | SDRL | MRL | ARL | SDRL | MRL | ARL | SDRL | MRL |
| Case 1: | 0.12 | 3.03 | 0.19 | 3 | 3.02 | 0.17 | 3 | 3.12 | 0.33 | 3 |
| $\phi_0 = 290$ | 0.14 | 3.65 | 0.61 | 4 | 3.58 | 0.60 | 4 | 3.98 | 0.63 | 4 |
| $\sigma_0 = 0.37$ | 0.16 | 5.38 | 1.44 | 5 | 5.25 | 1.39 | 5 | 6.10 | 1.67 | 6 |
| $\tau_0 = 155$ | 0.18 | 15.08 | 9.12 | 13 | 14.06 | 8.20 | 12 | 20.37 | 13.18 | 17 |
| | 0.20 | 373.87 | 363.74 | 264 | 310.78 | 308.19 | 215 | 891.57 | 850.61 | 627 |
| | 0.22 | 13.41 | 8.59 | 11 | 12.87 | 8.14 | 11 | 17 | 11.43 | 14 |
| | 0.24 | 5.48 | 2.10 | 5 | 5.37 | 2.03 | 5 | 6.11 | 2.40 | 6 |
| | 0.26 | 3.80 | 1.05 | 4 | 3.74 | 1.03 | 4 | 4.14 | 1.17 | 4 |
| | 0.28 | 3.11 | 0.72 | 3 | 3.06 | 0.71 | 3 | 3.31 | 0.75 | 3 |
| | LCL | 0.1775 | | | 0.1776 | | | 0.1753 | | |
| | UCL | 0.2225 | | | 0.2224 | | | 0.2247 | | |
| Case 2: | 0.12 | 3.66 | 0.56 | 4 | 3.74 | 0.56 | 4 | 3.74 | 0.55 | 4 |
| $\phi = 148$ | 0.14 | 4.76 | 0.97 | 5 | 4.88 | 1.01 | 5 | 4.86 | 1 | 5 |
| $\sigma_0 = 0.50$ | 0.16 | 7.83 | 2.89 | 7 | 8.10 | 3.07 | 7 | 8.11 | 3.03 | 7 |
| $\tau_0 = 96$ | 0.18 | 28.78 | 21.68 | 22 | 31.77 | 24.54 | 24 | 31.51 | 24.12 | 24 |
| | 0.20 | 371.43 | 370.29 | 258 | 437.76 | 426.40 | 307 | 434.45 | 429.41 | 303 |
| | 0.22 | 21.54 | 16.46 | 17 | 22.58 | 16.98 | 18 | 22.38 | 17.37 | 17 |
| | 0.24 | 7.72 | 3.86 | 7 | 8 | 4.04 | 7 | 7.92 | 3.98 | 7 |
| | 0.26 | 4.94 | 1.82 | 5 | 5.06 | 1.85 | 5 | 5.08 | 1.86 | 5 |
| | 0.28 | 3.83 | 1.14 | 4 | 3.92 | 1.17 | 4 | 3.93 | 1.17 | 4 |
| | LCL | 0.1696 | | | 0.1687 | | | 0.1570 | | |
| | UCL | 0.2304 | | | 0.2313 | | | 0.2430 | | |
| Case 3: | 0.12 | 5.05 | 0.94 | 5 | 5.07 | 0.93 | 5 | 5.05 | 0.95 | 5 |
| $\phi = 80$ | 0.14 | 7.22 | 2.17 | 7 | 7.23 | 2.19 | 7 | 7.24 | 2.22 | 7 |
| $\sigma_0 = 0.71$ | 0.16 | 14.54 | 7.88 | 13 | 14.69 | 7.95 | 13 | 14.53 | 7.92 | 12 |
| $\tau_0 = 51$ | 0.18 | 78.29 | 70.59 | 57 | 79.49 | 71.47 | 57 | 78.19 | 67.88 | 58 |
| | 0.20 | 371.57 | 365.40 | 260 | 377.76 | 370.57 | 265 | 373.99 | 363.78 | 264 |
| | 0.22 | 35.78 | 30.69 | 27 | 36.85 | 32.13 | 27 | 36.11 | 30.65 | 27 |
| | 0.24 | 12.12 | 7.86 | 10 | 12.17 | 7.94 | 10 | 12.17 | 7.91 | 10 |
| | 0.26 | 7.21 | 3.64 | 6 | 7.22 | 3.63 | 6 | 7.21 | 3.58 | 6 |
| | 0.28 | 5.27 | 2.15 | 5 | 5.30 | 2.16 | 5 | 5.26 | 2.13 | 5 |
| | LCL | 0.1571 | | | 0.1575 | | | 0.1569 | | |
| | UCL | 0.2429 | | | 0.2425 | | | 0.2431 | | |
| Case 4: | 0.12 | 10.69 | 3.88 | 10 | 9.62 | 3.32 | 9 | 9.85 | 3.46 | 9 |
| $\phi = 31$ | 0.14 | 21.26 | 13.02 | 18 | 17.91 | 10.20 | 15 | 18.60 | 10.98 | 15 |
| $\sigma_0 = 1.20$ | 0.16 | 74.03 | 62.49 | 56 | 53.09 | 44.06 | 40 | 58.90 | 49.99 | 43 |
| $\tau_0 = 20$ | 0.18 | 546.56 | 534.50 | 385 | 305.40 | 294.95 | 215 | 359.71 | 349.53 | 256 |
| | 0.20 | 370.23 | 359.05 | 262 | 270.13 | 261.48 | 192 | 296.85 | 295.46 | 205 |
| | 0.22 | 67.19 | 62.48 | 48 | 57.21 | 52.33 | 41 | 59.71 | 54.78 | 43 |
| | 0.24 | 24.83 | 20.13 | 19 | 21.88 | 17.75 | 17 | 22.67 | 18.24 | 17 |
| | 0.26 | 13.68 | 9.54 | 11 | 12.44 | 8.58 | 10 | 12.81 | 8.87 | 10 |
| | 0.28 | 9.26 | 5.56 | 8 | 8.67 | 5.19 | 7 | 8.81 | 5.25 | 7 |
| | LCL | 0.1275 | | | 0.1320 | | | 0.1310 | | |
| | UCL | 0.2725 | | | 0.2680 | | | 0.2690 | | |

**Table 13:** Performance of two-sided EWMA Charts, $\lambda = 0.05$, True Model: Unit Gamma distribution.

| | | Unit Gamma - True | | | Beta | | | Simplex | | |
|---|---|---|---|---|---|---|---|---|---|---|
| | $\mu$ | ARL | SDRL | MRL | ARL | SDRL | MRL | ARL | SDRL | MRL |
| Case 1: | 0.12 | 4.23 | 0.52 | 4 | 3.87 | 0.52 | 4 | 3.88 | 0.52 | 4 |
| $\phi_0 = 290$ | 0.14 | 5.26 | 0.85 | 5 | 4.74 | 0.79 | 5 | 4.74 | 0.79 | 5 |
| $\sigma_0 = 0.37$ | 0.16 | 7.49 | 1.73 | 7 | 6.66 | 1.58 | 6 | 6.66 | 1.57 | 6 |
| $\tau_0 = 155$ | 0.18 | 15.74 | 6.29 | 14 | 13.27 | 5.34 | 12 | 13.34 | 5.31 | 12 |
| | 0.20 | 370.39 | 356.35 | 260 | 186.66 | 175.29 | 133 | 182.72 | 171.12 | 130 |
| | 0.22 | 15.91 | 6.86 | 15 | 13.46 | 5.86 | 12 | 13.32 | 5.70 | 12 |
| | 0.24 | 7.60 | 2.01 | 7 | 6.70 | 1.81 | 6 | 6.73 | 1.80 | 6 |
| | 0.26 | 5.29 | 1.04 | 5 | 4.79 | 0.96 | 5 | 4.81 | 1 | 5 |
| | 0.28 | 4.26 | 0.70 | 4 | 3.90 | 0.68 | 4 | 3.91 | 0.68 | 4 |
| | LCL | | 0.1897 | | | 0.1907 | | | 0.1906 | |
| | UCL | | 0.2103 | | | 0.2093 | | | 0.2094 | |
| Case 2: | 0.12 | 5.05 | 0.75 | 5 | 5.01 | 0.75 | 5 | 5.01 | 0.74 | 5 |
| $\phi = 148$ | 0.14 | 6.43 | 1.25 | 6 | 6.40 | 1.23 | 6 | 6.37 | 1.22 | 6 |
| $\sigma_0 = 0.50$ | 0.16 | 9.44 | 2.60 | 9 | 9.40 | 2.60 | 9 | 9.39 | 2.57 | 9 |
| $\tau_0 = 96$ | 0.18 | 20.95 | 9.51 | 19 | 21.05 | 9.93 | 19 | 20.93 | 9.80 | 19 |
| | 0.20 | 370.50 | 351.26 | 262 | 360.00 | 346.57 | 252 | 364.01 | 353.78 | 253 |
| | 0.22 | 21.44 | 10.71 | 19 | 21.00 | 10.67 | 19 | 21.04 | 10.61 | 19 |
| | 0.24 | 9.54 | 3.01 | 9 | 9.50 | 3 | 9 | 9.52 | 2.97 | 9 |
| | 0.26 | 6.50 | 1.54 | 6 | 6.47 | 1.54 | 6 | 6.46 | 1.54 | 6 |
| | 0.28 | 5.12 | 1.01 | 5 | 5.06 | 1 | 5 | 5.09 | 0.99 | 5 |
| | LCL | | 0.1869 | | | 0.1870 | | | 0.1874 | |
| | UCL | | 0.2131 | | | 0.2130 | | | 0.2126 | |
| Case 3: | 0.12 | 6.56 | 1.20 | 6 | 6.58 | 1.20 | 6 | 6.57 | 1.22 | 6 |
| $\phi = 80$ | 0.14 | 8.60 | 2.11 | 8 | 8.67 | 2.09 | 8 | 8.63 | 2.09 | 8 |
| $\sigma_0 = 0.71$ | 0.16 | 13.27 | 4.50 | 12 | 13.35 | 4.55 | 12 | 13.34 | 4.65 | 12 |
| $\tau_0 = 51$ | 0.18 | 32.33 | 18.66 | 28 | 32.79 | 19.09 | 28 | 32.80 | 19.13 | 28 |
| | 0.20 | 370.22 | 358.01 | 260 | 387.68 | 376.96 | 273 | 381.90 | 377.61 | 263 |
| | 0.22 | 32.10 | 19.59 | 27 | 32.02 | 19.52 | 27 | 32.40 | 19.71 | 27 |
| | 0.24 | 13.46 | 5.35 | 12 | 13.48 | 5.31 | 13 | 13.54 | 5.39 | 12 |
| | 0.26 | 8.75 | 2.61 | 8 | 8.79 | 2.65 | 8 | 8.77 | 2.62 | 8 |
| | 0.28 | 6.64 | 1.63 | 6 | 6.70 | 1.67 | 6 | 6.66 | 1.63 | 6 |
| | LCL | | 0.1821 | | | 0.1823 | | | 0.1822 | |
| | UCL | | 0.2179 | | | 0.2177 | | | 0.2178 | |
| Case 4: | 0.12 | 10.22 | 2.59 | 10 | 10.06 | 2.55 | 10 | 10.46 | 2.68 | 10 |
| $\phi = 31$ | 0.14 | 14.10 | 4.73 | 13 | 13.76 | 4.62 | 13 | 14.47 | 4.83 | 13 |
| $\sigma_0 = 1.20$ | 0.16 | 23.67 | 11.22 | 21 | 23.04 | 11.02 | 20 | 24.29 | 11.61 | 21 |
| $\tau_0 = 20$ | 0.18 | 65.20 | 47.64 | 52 | 63.80 | 45.82 | 51 | 69.02 | 50.26 | 55 |
| | 0.20 | 370.05 | 352.40 | 261 | 341.50 | 323.28 | 246 | 415.34 | 403.81 | 294 |
| | 0.22 | 60.87 | 46.88 | 48 | 59.10 | 44.92 | 46 | 63.70 | 48.76 | 50 |
| | 0.24 | 23.67 | 12.78 | 21 | 23.20 | 12.62 | 20 | 24.23 | 12.94 | 21 |
| | 0.26 | 14.31 | 5.95 | 13 | 14.14 | 5.84 | 13 | 14.61 | 6.09 | 13 |
| | 0.28 | 10.46 | 3.66 | 10 | 10.26 | 3.58 | 10 | 10.68 | 3.73 | 10 |
| | LCL | | 0.1716 | | | 0.1719 | | | 0.1704 | |
| | UCL | | 0.2284 | | | 0.2281 | | | 0.2296 | |

**Table 14:** Performance of two-sided EWMA Charts, $\lambda = 0.10$, True Model: Unit Gamma distribution.

| | | Unit Gamma - True | | | Beta | | | Simplex | | |
|---|---|---|---|---|---|---|---|---|---|---|
| | $\mu$ | ARL | SDRL | MRL | ARL | SDRL | MRL | ARL | SDRL | MRL |
| Case 1: | 0.12 | 3.65 | 0.57 | 4 | 3.48 | 0.55 | 3 | 3.48 | 0.54 | 3 |
| $\phi_0 = 290$ | 0.14 | 4.53 | 0.81 | 4 | 4.33 | 0.79 | 4 | 4.31 | 0.79 | 4 |
| $\sigma_0 = 0.37$ | 0.16 | 6.54 | 1.76 | 6 | 6.13 | 1.65 | 6 | 6.16 | 1.65 | 6 |
| $\tau_0 = 155$ | 0.18 | 15.09 | 7.30 | 13 | 13.65 | 6.72 | 12 | 13.63 | 6.67 | 12 |
| | 0.20 | 370.58 | 360.56 | 259 | 242.05 | 234.89 | 169 | 242.03 | 234.51 | 171 |
| | 0.22 | 15.10 | 7.94 | 13 | 13.65 | 7 | 12 | 13.58 | 6.96 | 12 |
| | 0.24 | 6.63 | 2.03 | 6 | 6.25 | 1.92 | 6 | 6.20 | 1.90 | 6 |
| | 0.26 | 4.60 | 1.02 | 4 | 4.37 | 0.97 | 4 | 4.40 | 0.99 | 4 |
| | 0.28 | 3.71 | 0.68 | 4 | 3.55 | 0.65 | 3 | 3.55 | 0.64 | 3 |
| | LCL | | 0.1840 | | | 0.1855 | | | 0.1854 | |
| | UCL | | 0.2160 | | | 0.2145 | | | 0.2146 | |
| Case 2: | 0.12 | 4.35 | 0.70 | 4 | 4.32 | 0.69 | 4 | 4.31 | 0.70 | 4 |
| $\phi = 148$ | 0.14 | 5.58 | 1.24 | 5 | 5.50 | 1.22 | 5 | 5.51 | 1.22 | 5 |
| $\sigma_0 = 0.50$ | 0.16 | 8.39 | 2.72 | 8 | 8.26 | 2.70 | 8 | 8.25 | 2.71 | 8 |
| $\tau_0 = 96$ | 0.18 | 21.59 | 12.65 | 18 | 20.84 | 12.38 | 18 | 20.78 | 12.17 | 18 |
| | 0.20 | 370.30 | 368.98 | 256 | 334.45 | 332.52 | 234 | 330.71 | 319.12 | 234 |
| | 0.22 | 20.94 | 12.86 | 18 | 20.74 | 12.94 | 17 | 20.28 | 12.44 | 17 |
| | 0.24 | 8.52 | 3.23 | 8 | 8.41 | 3.18 | 8 | 8.44 | 3.16 | 8 |
| | 0.26 | 5.64 | 1.55 | 5 | 5.55 | 1.52 | 5 | 5.57 | 1.51 | 5 |
| | 0.28 | 4.42 | 0.96 | 4 | 4.37 | 0.95 | 4 | 4.36 | 0.95 | 4 |
| | LCL | | 0.1797 | | | 0.1898 | | | 0.1803 | |
| | UCL | | 0.2203 | | | 0.2202 | | | 0.2197 | |
| Case 3: | 0.12 | 5.67 | 1.19 | 5 | 5.70 | 1.21 | 6 | 5.73 | 1.21 | 6 |
| $\phi = 80$ | 0.14 | 7.57 | 2.16 | 7 | 7.62 | 2.18 | 7 | 7.63 | 2.18 | 7 |
| $\sigma_0 = 0.71$ | 0.16 | 12.41 | 5.22 | 11 | 12.54 | 5.27 | 11 | 12.52 | 5.32 | 11 |
| $\tau_0 = 51$ | 0.18 | 37.36 | 27 | 30 | 37.22 | 26.85 | 29 | 37.58 | 26.97 | 30 |
| | 0.20 | 370.64 | 364.28 | 260 | 383.61 | 376.51 | 268 | 379.20 | 373.13 | 264 |
| | 0.22 | 34.10 | 24.84 | 27 | 34.84 | 25.70 | 27 | 34.78 | 25.62 | 27 |
| | 0.24 | 12.48 | 6 | 11 | 12.51 | 5.94 | 11 | 12.53 | 6.01 | 11 |
| | 0.26 | 7.73 | 2.72 | 7 | 7.72 | 2.71 | 7 | 7.80 | 2.76 | 7 |
| | 0.28 | 5.80 | 1.64 | 6 | 5.83 | 1.65 | 6 | 5.81 | 1.64 | 6 |
| | LCL | | 0.1722 | | | 0.1725 | | | 0.1723 | |
| | UCL | | 0.2278 | | | 0.2275 | | | 0.2277 | |
| Case 4: | 0.12 | 9.22 | 2.82 | 9 | 9.19 | 2.87 | 9 | 9.72 | 3.03 | 9 |
| $\phi = 31$ | 0.14 | 13.56 | 5.77 | 12 | 13.32 | 5.65 | 12 | 14.18 | 6.02 | 13 |
| $\sigma_0 = 1.20$ | 0.16 | 25.25 | 15.16 | 21 | 25.08 | 15.20 | 21 | 28.02 | 17.62 | 23 |
| $\tau_0 = 20$ | 0.18 | 92.48 | 80.67 | 68 | 89.51 | 77.34 | 67 | 111.80 | 99.18 | 82 |
| | 0.20 | 370.49 | 362.90 | 258 | 354.58 | 348.29 | 247 | 484.07 | 481.67 | 335 |
| | 0.22 | 67.42 | 57.75 | 50 | 66.53 | 57.10 | 49 | 77.14 | 67.27 | 57 |
| | 0.24 | 23.59 | 15.48 | 19 | 23.52 | 15.70 | 19 | 25.61 | 17.28 | 21 |
| | 0.26 | 13.35 | 6.78 | 12 | 13.48 | 6.83 | 12 | 14.26 | 7.35 | 13 |
| | 0.28 | 9.43 | 3.86 | 9 | 9.38 | 3.87 | 9 | 9.95 | 4.20 | 9 |
| | LCL | | 0.1557 | | | 0.1562 | | | 0.1539 | |
| | UCL | | 0.2443 | | | 0.2438 | | | 0.2461 | |

**Table 15:** Performance of two-sided EWMA Charts, $\lambda = 0.20$, True Model: Unit Gamma distribution.

| | | Unit Gamma - True | | | Beta | | | Simplex | | |
|---|---|---|---|---|---|---|---|---|---|---|
| | $\mu$ | ARL | SDRL | MRL | ARL | SDRL | MRL | ARL | SDRL | MRL |
| Case 1: | 0.12 | 3.21 | 0.45 | 3 | 3.04 | 0.41 | 3 | 3.04 | 0.41 | 3 |
| $\phi_0 = 290$ | 0.14 | 3.96 | 0.84 | 4 | 3.64 | 0.77 | 4 | 3.65 | 0.76 | 4 |
| $\sigma_0 = 0.37$ | 0.16 | 6.01 | 2.01 | 6 | 5.28 | 1.73 | 5 | 5.29 | 1.73 | 5 |
| $\tau_0 = 155$ | 0.18 | 16.94 | 11.06 | 14 | 12.99 | 7.94 | 11 | 13.11 | 8.23 | 11 |
| | 0.20 | 370.54 | 369.01 | 258 | 165.74 | 160.42 | 117 | 164.62 | 159.83 | 116 |
| | 0.22 | 16.01 | 10.76 | 13 | 12.68 | 8.03 | 10 | 12.74 | 8.18 | 11 |
| | 0.24 | 6.02 | 2.28 | 6 | 5.35 | 1.97 | 5 | 5.37 | 1.98 | 5 |
| | 0.26 | 4.06 | 1.04 | 4 | 3.72 | 0.93 | 4 | 3.70 | 0.93 | 4 |
| | 0.28 | 3.26 | 0.64 | 3 | 3.05 | 0.61 | 3 | 3.04 | 0.61 | 3 |
| | $LCL$ | 0.1753 | | | 0.1776 | | | 0.1775 | | |
| | $UCL$ | 0.2247 | | | 0.2224 | | | 0.2225 | | |
| Case 2: | 0.12 | 3.79 | 0.74 | 4 | 3.76 | 0.73 | 4 | 3.66 | 0.71 | 4 |
| $\phi = 148$ | 0.14 | 4.99 | 1.36 | 5 | 4.89 | 1.29 | 5 | 4.75 | 1.23 | 5 |
| $\sigma_0 = 0.50$ | 0.16 | 8.22 | 3.45 | 7 | 7.98 | 3.35 | 7 | 7.56 | 3.15 | 7 |
| $\tau_0 = 96$ | 0.18 | 27.55 | 20.99 | 21 | 26.69 | 20.02 | 21 | 23.99 | 17.46 | 19 |
| | 0.20 | 370.97 | 369.27 | 256 | 340.51 | 333.03 | 237 | 262.56 | 256.87 | 182 |
| | 0.22 | 24.58 | 19.58 | 19 | 23.68 | 18.46 | 18 | 21.43 | 16.11 | 17 |
| | 0.24 | 8.21 | 3.94 | 7 | 7.99 | 3.76 | 7 | 7.65 | 3.64 | 7 |
| | 0.26 | 5.11 | 1.69 | 5 | 4.99 | 1.62 | 5 | 4.83 | 1.55 | 5 |
| | 0.28 | 3.92 | 0.98 | 4 | 3.87 | 0.97 | 4 | 3.75 | 0.92 | 4 |
| | $LCL$ | 0.1570 | | | 0.1687 | | | 0.1696 | | |
| | $UCL$ | 0.2430 | | | 0.2313 | | | 0.2304 | | |
| Case 3: | 0.12 | 5.10 | 1.33 | 5 | 5.11 | 1.29 | 5 | 5.10 | 1.31 | 5 |
| $\phi = 80$ | 0.14 | 7.21 | 2.70 | 7 | 7.18 | 2.59 | 7 | 7.17 | 2.58 | 7 |
| $\sigma_0 = 0.71$ | 0.16 | 13.56 | 7.71 | 12 | 13.35 | 7.61 | 11 | 13.60 | 7.78 | 12 |
| $\tau_0 = 51$ | 0.18 | 58.06 | 50.63 | 43 | 57.61 | 50.55 | 42 | 57.95 | 50.72 | 43 |
| | 0.20 | 370.62 | 364.79 | 261 | 366.55 | 355.93 | 259 | 365.66 | 358.82 | 257 |
| | 0.22 | 41.43 | 35.01 | 31 | 41.65 | 36.62 | 30 | 41.27 | 35.22 | 31 |
| | 0.24 | 12.90 | 8 | 11 | 12.86 | 8 | 11 | 12.97 | 8.02 | 11 |
| | 0.26 | 7.28 | 3.21 | 7 | 7.28 | 3.24 | 7 | 7.21 | 3.20 | 7 |
| | 0.28 | 5.24 | 1.78 | 5 | 5.22 | 1.77 | 5 | 5.22 | 1.77 | 5 |
| | $LCL$ | 0.1569 | | | 0.1575 | | | 0.1571 | | |
| | $UCL$ | 0.2431 | | | 0.2425 | | | 0.2429 | | |
| Case 4: | 0.12 | 9.47 | 3.98 | 8 | 9.19 | 3.82 | 8 | 9.91 | 4.21 | 9 |
| $\phi = 31$ | 0.14 | 15.87 | 9.37 | 13 | 15.49 | 8.94 | 13 | 17.23 | 10.51 | 14 |
| $\sigma_0 = 1.20$ | 0.16 | 40.62 | 32.82 | 31 | 37.62 | 29.59 | 29 | 47.98 | 40.20 | 36 |
| $\tau_0 = 20$ | 0.18 | 219.01 | 212.44 | 154 | 192.60 | 187.68 | 134 | 286.12 | 277.19 | 200 |
| | 0.20 | 370.41 | 371.31 | 256 | 330.26 | 326.91 | 231 | 459.47 | 461.78 | 317 |
| | 0.22 | 79.77 | 74.21 | 57 | 75.79 | 70.80 | 54 | 91.58 | 86.22 | 65 |
| | 0.24 | 27.75 | 21.93 | 21 | 26.58 | 21.18 | 21 | 30.05 | 24.43 | 23 |
| | 0.26 | 14.13 | 9.25 | 12 | 13.77 | 9.03 | 11 | 15.09 | 9.85 | 12 |
| | 0.28 | 9.34 | 4.93 | 8 | 9.09 | 4.73 | 8 | 9.75 | 5.19 | 8 |
| | $LCL$ | 0.1310 | | | 0.1320 | | | 0.1275 | | |
| | $UCL$ | 0.2690 | | | 0.2680 | | | 0.2725 | | |

## 5. Numerical Example

In this section we present a real data example for the practical application of the considered EWMA charts. We use the data set from Sant'Anna and ten Caten (2012) (see also Ho et al. (2019)) about the proportion of contaminated peanuts. 34 individual observations are available (see Table A2 in the Appendix), and each value gives the proportion of non-contaminated peanuts in 34 batches of 120 pounds each. We consider this data set as a benchmark one due to its popularity in the related studies. Observations 1-20 are used as a Phase I sample.

First, we apply the runs test for randomness (Gibbons and Chakraborti (2021), p. 75) to test the hypothesis that the observations constitute a random sample. The p-value equals 0.3581 which indicates that the available data are realizations from independent and identically distributed r.v. Then, by using the method of maximum likelihood estimation, we provide in Table 16 the point estimates (and the estimated standard errors, in the parentheses) of the parameters for the three candidate models. The results verify those in Ho et al. (2019). The function `optim` of R (R Core Team (2022)) with the argument `method = 'BFGS'` (which implements the Quasi-Newton optimization method BFGS. See, for example, Fletcher (2013)) has been used for estimating the parameters of each model.

Also, we provide the values of model selection criteria AIC (Akaike's Information Criterion) and BIC (Bayesian Information Criterion), in the respective columns as well as the values of the test statistics and the respective p-values for the Anderson-Darling and Kolmogorov-Smirnov goodness-of-fit tests. See columns 'AD, pvalue' and 'KS, pvalue', respectively. Specifically, both the Beta and the Unit Gamma models have the same fit in the data while the Simplex model is slightly better. Moreover, we provide in Figure 2 the empirical distribution function (e.c.d.f.) of the Phase I data along with the c.d.f. of each model. Thus, from the above analysis and from the comparison between the values of the AIC, BIC among the different models as well as the p-values of the goodness-of-fit tests we conclude that the Simplex distribution is the model with the best fit in the data.

**Table 16**: Estimates of model parameters and values of model selection criteria.

| Model | Estimates | AIC BIC | AD, pvalue | KS, pvalue |
|---|---|---|---|---|
| Beta | $\hat{\mu} = 0.9533\ (0.00667)$ $\hat{\phi} = 48.9438\ (15.95920)$ | -85.455 -83.464 | 0.4970, 0.7478 | 0.1624, 0.6684 |
| Simplex | $\hat{\mu} = 0.9534\ (0.00718)$ $\hat{\sigma} = 3.5742\ (0.56498)$ | -88.653 -86.662 | 0.2397, 0.9755 | 0.1310, 0.8825 |
| Unit Gamma | $\hat{\mu} = 0.9534\ (0.00666)$ $\hat{\tau} = 2.2798\ (0.67487)$ | -85.455 -83.463 | 0.4966, 0.7482 | 0.1603, 0.6805 |

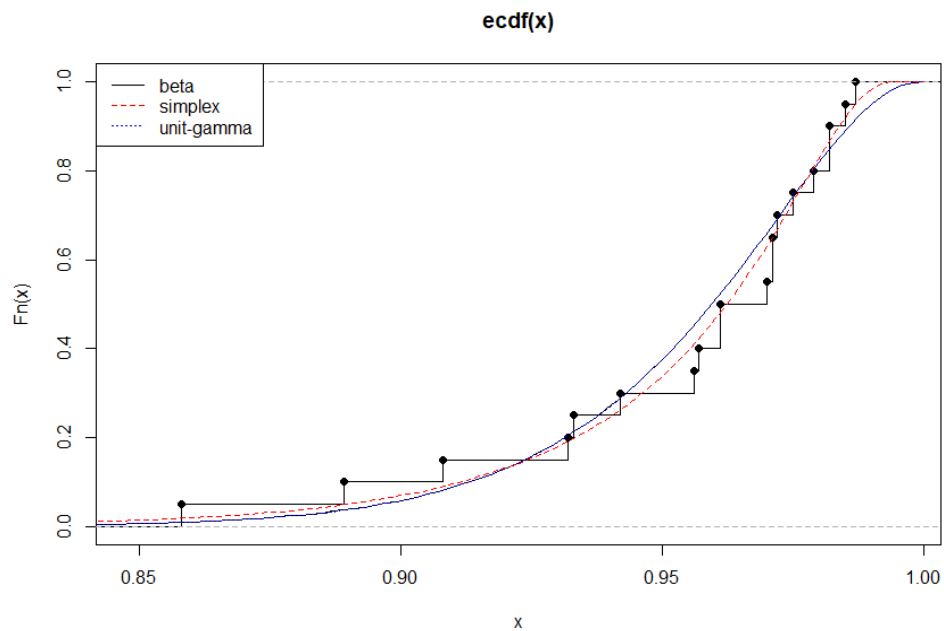

**Figure 2:** Empirical c.d.f. for the Phase I data.

Next, we will consider only the Simplex distribution and we will construct the two-sided Shewhart and EWMA control charts that are based on this model.

First, we apply the two-sided Shewhart chart as a Phase I data method to verify that the process was IC, when the data have been collected. For illustrative purposes we choose a false alarm rate equal to 0.0027 and therefore, the desired $ARL_0 = 370.4$. The control limits of the two-sided Shewhart chart are calculated as the 0.00135 and 0.99865 percentile points of the Simplex distribution with parameters $\mu = 0.9534$ and $\sigma = 3.5742$. Their values are $LCL = 0.7794$, $UCL = 0.9936$. The $CL$ is placed at 0.9534, the IC process mean level. From Figure 3, all the points are within the control limits and thus, the process was IC when the data have been collected. This was also mentioned in the work of Ho et al. (2019).

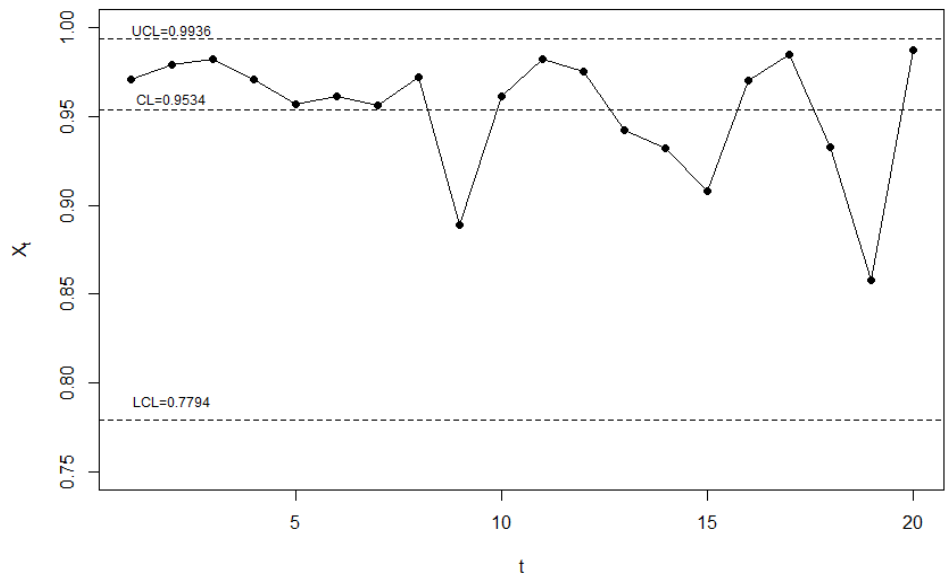

**Figure 3:** Phase I Shewhart Chart (Simplex model) for the proportions of non-contaminated peanuts.

Then we proceed to the Phase II analysis by assuming that the estimates of parameters $\mu$, $\sigma$ are the true IC parameter values. In Figure 4 we provide the two-sided Simplex Shewhart and EWMA charts for the 14 remaining points (21-34), which are considered as the Phase II observations. For the two-sided EWMA chart we applied the steps of the algorithmic procedure (see Table 1) and determined the control limits for $\lambda \in \{0.05, 0.10, 0.20\}$. The Shewhart chart gives an OOC signal for the first time in point 12 while the EWMA charts signal at observation 5 (for $\lambda = 0.05$ or $0.10$) or at observation 4 (for $\lambda = 0.20$). This is an indication of process deterioration since the proportion of non-contaminated peanuts decreases. Compared to the Shewhart chart, the EWMA chart detects this shift about 7-8 points earlier. This fact also highlights its importance in quick detection of shifts in process mean level.

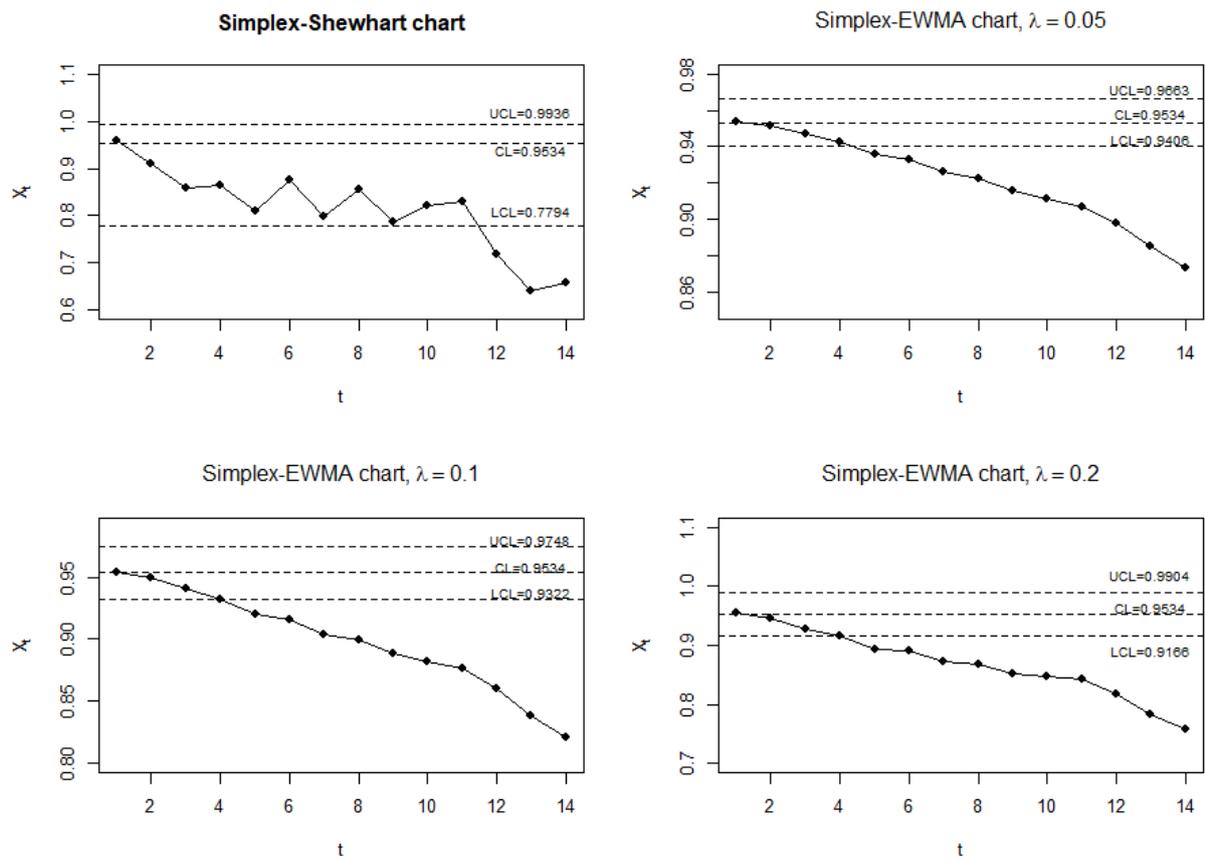

**Figure 4:** Phase II two-sided Simplex Shewhart and EWMA charts.

## 6. Conclusions and Future Research

In this work we studied the performance of two-sided EWMA charts for individual observations in the monitoring of double bounded processes and specifically, in the interval (0,1). Three different probability models have been considered as candidate models. Then, two-sided EWMA-type control charts have been designed for three different values for the smoothing parameter $\lambda$, for each of the three models. The performance of the proposed charts was studied via simulation and compared to the performance of the Shewhart-type charts (for each of the three models), which have been studied by Ho et al. (2019). The results showed the superiority of the proposed EWMA charts; for specific cases, the EWMA charts outperform the Shewhart chart even for large shifts in process mean average. The improvement in the detection of small shifts (e.g. in the interval [0.18, 0.22]) is from 70% up to 90%.

In addition, we investigated how much the EWMA charts are affected when control limits that have been established under a 'wrong' model are used for monitoring a process that it is modeled by the 'true' model. Our extensive simulation study revealed that in almost all the considered scenarios (which include distributions with large and small variances, close to symmetry or asymmetric ones), the IC performance is mostly affected. Moreover, in cases of large process variance, the OOC performance of the chart is affected, especially for small shifts in process mean level. However, for moderate to large shifts, the use of 'wrong' limits does not have a significant effect on the OOC *ARL* values, between the three different models.

In practical problems, the best approach is to conduct first a detailed model selection procedure among the possible candidate models and choose the one with the best fit in the data. Then, we advise practitioners to apply the two-sided EWMA chart with $\lambda = 0.05$ and $L$ around 2.49 or with $\lambda = 0.10$ and $L$ around 2.70, whichever is the true probability model for the continuous proportions. For all the considered cases, these charts have better performance than the Shewhart chart in small increasing and decreasing shifts, whereas their out-of-control performance is not very much affected, whether the correct limits are used or not. In addition, these $(\lambda, L)$ pairs reduce the *ARL*-biased performance of the EWMA chart.

Possible topics for future research consist of a study where both process parameters can change, as well as a study on the performance of charts when process parameters are unknown and need to be estimated. Moreover, the present study can be extended by considering other probability models such as the Kumaraswamy and the unit-Weibull distributions, as well as to CUSUM-type charts.

Finally, the codes for reproducing the results in this work have been prepared in R and they are available from the authors upon request.

**Appendix**

Table A.1: $L$ values for each model for $\lambda \in \{0.05, 0.10, 0.20\}$ and $ARL_0 = 370.4$.

| Beta | | | Simplex | | | Unit Gamma | | |
|---|---|---|---|---|---|---|---|---|
| $\phi$ | $\lambda$ | $L$ | $\sigma$ | $\lambda$ | $L$ | $\tau$ | $\lambda$ | $L$ |
| 290 | 0.05 | 2.481 | 0.37 | 0.05 | 2.491 | 155 | 0.05 | 2.492 |
|  | 0.10 | 2.701 |  | 0.10 | 2.700 |  | 0.10 | 2.703 |
|  | 0.20 | 2.861 |  | 0.20 | 2.866 |  | 0.20 | 2.864 |
| 148 | 0.05 | 2.485 | 0.5 | 0.05 | 2.491 | 96 | 0.05 | 2.497 |
|  | 0.10 | 2.693 |  | 0.10 | 2.705 |  | 0.10 | 2.701 |
|  | 0.20 | 2.864 |  | 0.20 | 2.874 |  | 0.20 | 2.872 |
| 80 | 0.05 | 2.487 | 0.71 | 0.05 | 2.489 | 51 | 0.05 | 2.491 |
|  | 0.10 | 2.701 |  | 0.10 | 2.703 |  | 0.10 | 2.697 |
|  | 0.20 | 2.869 |  | 0.20 | 2.882 |  | 0.20 | 2.875 |
| 31 | 0.05 | 2.483 | 1.2 | 0.05 | 2.528 | 20 | 0.05 | 2.487 |
|  | 0.10 | 2.702 |  | 0.10 | 2.752 |  | 0.10 | 2.704 |
|  | 0.20 | 2.884 |  | 0.20 | 2.977 |  | 0.20 | 2.899 |

Table A.2: Proportions of contaminated peanuts

| $t$ | 1 | 2 | 3 | 4 | 5 | 6 | 7 | 8 | 9 | 10 |
|---|---|---|---|---|---|---|---|---|---|---|
| $X_t$ | 0.971 | 0.979 | 0.982 | 0.971 | 0.957 | 0.961 | 0.956 | 0.972 | 0.889 | 0.961 |
| $t$ | 11 | 12 | 13 | 14 | 15 | 16 | 17 | 18 | 19 | 20 |
| $X_t$ | 0.982 | 0.975 | 0.942 | 0.932 | 0.908 | 0.970 | 0.985 | 0.933 | 0.858 | 0.987 |
| $t$ | 21 | 22 | 23 | 24 | 25 | 26 | 27 | 28 | 29 | 30 |
| $X_t$ | 0.958 | 0.909 | 0.859 | 0.863 | 0.811 | 0.877 | 0.798 | 0.855 | 0.788 | 0.821 |
| $t$ | 31 | 32 | 33 | 34 | | | | | | |
| $X_t$ | 0.830 | 0.718 | 0.642 | 0.658 | | | | | | |